\documentclass[useAMS,useusegraphicx]{mn2e}

\usepackage{graphicx}
\usepackage{amsmath}
\usepackage{mathtools}
\usepackage{amssymb}
\usepackage{wasysym}
\usepackage{txfonts}

\usepackage{latexsym,longtable,epsf,ulem}
\usepackage{cases} 
\usepackage{enumerate} 

\newcommand{\ant}{\alpha_{\rm nt}}
\newcommand{\hii}{H{\sc ii} }

\newcommand{\mnras}{MNRAS}
\newcommand{\araa}{ARA\&A}
\newcommand{\aj}{AJ}
\newcommand{\apj}{ApJ}

\newcommand{\apjs}{ApJS}

\newcommand{\aap}{A\&A}

\newcommand{\pasa}{PASA}

\newcommand{\physrep}{Phys.~Rep.}

\title[Spectral index and the ISM]
{Synchrotron spectral index and interstellar medium densities of star-forming galaxies}
\author[Basu et al.]{Aritra Basu$^1$\thanks{E-mail : abasu@mpifr-bonn.mpg.de}, 
	Rainer Beck$^1$, 
	Philip Schmidt$^1$, 
	Subhashis Roy$^2$\\
$^1$Max-Planck-Institut f{\"u}r Radioastronomie, Auf dem H{\"u}gel 69, D-53121 Bonn, Germany\\
$^2$National Centre for Radio Astrophysics, TIFR, Pune University Campus, Ganeshkhind Road, Pune -- 411007, India\\}

\begin{document}


\pagerange{\pageref{firstpage}--\pageref{lastpage}} \pubyear{2002}

\maketitle

\label{firstpage}

\begin{abstract}

The spectral index of synchrotron emission is an important parameter in
understanding the properties of cosmic ray electrons (CREs) and the
interstellar medium (ISM). We determine the synchrotron spectral index ($\ant$)
of four nearby star-forming galaxies, namely NGC 4736, NGC 5055, NGC 5236 and
NGC 6946 at sub-kpc linear scales.  The $\ant$ was determined between 0.33 and
1.4 GHz for all the galaxies.  We find the spectral index to be flatter
($\gtrsim -0.7$) in regions with total neutral (atomic + molecular) gas surface
density, $\Sigma_{\rm gas} \gtrsim \rm 50~M_\odot pc^{-2}$,
typically in the arms and inner parts of the galaxies. In regions with
$\Sigma_{\rm gas} \lesssim \rm 50~M_\odot pc^{-2}$, especially in the interarm
and outer regions of the galaxies, the spectral index steepens sharply to
$<-1.0$.  The flattening of $\ant$ is unlikely to be caused due to thermal
free--free absorption at 0.33 GHz.  Our result is consistent with the scenario
where the CREs emitting at frequencies below $\sim0.3$ GHz are dominated by
bremsstrahlung and/or ionization losses. For denser medium ($\Sigma_{\rm gas}
\gtrsim \rm 200~M_\odot pc^{-2}$), having strong magnetic fields ($\sim
30~\mu$G), $\ant$ is seen to be flatter than $-0.5$, perhaps caused due to
ionization losses. We find that, due to the clumpy nature of the ISM,
such dense regions cover only a small fraction of the galaxy ($\lesssim5$
percent). Thus, the galaxy-integrated spectrum may not show indication of such
loss mechanisms and remain a power-law over a wide range of radio frequencies
(between $\sim 0.1$ to 10 GHz).

\end{abstract}

\begin{keywords}
	galaxies : ISM -- (ISM:) cosmic rays -- ISM : magnetic fields -- galaxies: ISM 
	-- radiation mechanisms : non-thermal -- radio continuum : galaxies
\end{keywords}

\section{Introduction}

The synchrotron spectral index of the cosmic ray electrons (CREs) in normal
star-forming galaxies is an important parameter in understanding their energy
loss/gain mechanisms in the interstellar medium (ISM). Studying spatially
resolved properties of the synchrotron (also referred to as non-thermal)
spectral index\footnote{Spectral index, $\alpha$, is defined as: $S_\nu \propto
\nu^{\alpha}$, where $S_\nu$ is the radio continuum flux density at a frequency
$\nu$.} ($\ant$) can throw meaningful insights into the properties of the ISM
and thereby understand the nature of multi waveband emission from galaxies. 

It is believed that the CREs are produced by diffusive shock acceleration (DSA)
of thermal electrons up to energies of few PeV in shock fronts of supernova
remnants \nocite{bell04, jones11, edmon11, kang12}({Bell} 2004; {Jones} 2011; {Edmon} {et~al.} 2011; {Kang}, {Edmon} \& {Jones} 2012). The CREs are injected into
the ISM with a power-law energy spectrum $n(E)dE \propto E^{\gamma}$, where,
$n(E)$ is the number density of CREs with energies in the range $E$ and $E+dE$.
The energy spectral index, $\gamma$, of CREs is related to $\ant$ as $\gamma =
2\ant -1$ and can be measured directly from observations.  Depending on the CR
acceleration mechanism and efficiency, $\gamma$ is estimated to lie in the
range $-2$ and $-2.4$ \nocite{bierm93, bogda83, bland87, bell78}({Biermann} \& {Strom} 1993; {Bogdan} \& {V{\"o}lk} 1983; {Blandford} \& {Eichler} 1987; {Bell} 1978), which
corresponds to $\ant$ in the range $-0.5$ and $-0.7$.

The CREs that are injected into the ISM are subjected to energy dependent
losses and hence $\ant$ is expected to be modified smoothly over a frequency
range. The main energy loss mechanisms involving ionization losses,
relativistic bremsstrahlung losses, synchrotron and inverse-Compton losses
alter the radio frequency spectrum in a specific way.  These loss
mechanisms have different dependence on CRE energy \nocite{longa11}(see {Longair} 2011) and
thus they are important in different parts of the spectrum.
Ionization and relativistic bremsstrahlung losses typically affect CREs 
having energies $\lesssim1$ GeV while synchrotron and inverse-Compton losses
affect CREs having energies $\gtrsim2$ GeV.  These energies correspond to
$\sim 0.2$ GHz and $\sim1$ GHz, respectively, for typical magnetic fields of
$\sim 10~\mu$G. 

At higher radio frequencies, synchrotron and inverse-Compton losses lead to a
cut-off in the spectrum.  However, the form of this cut-off depends on the
pitch angle between the magnetic field and CREs and the process of particle
injection.  In the case of steady continuous injection (CI) of CREs, the
spectrum steepens by 0.5, i.e., ($\Delta \alpha = -0.5$) \nocite{pacho70}({Pacholczyk} 1970).  For
injection of particles at a single epoch and considering constant pitch angles
of individual electrons, according to the Kardashev-Pacholczyk model
\nocite{karda62, pacho70}(KP; {Kardashev} 1962; {Pacholczyk} 1970) the spectrum falls off as a power-law
with index $4\alpha_{\rm inj}/3 -1$, i.e., $\Delta \alpha = \alpha_{\rm inj}/3
-1$. Here, $\alpha_{\rm inj}$ is the initial power-law index of the particles.
The Jaffe-Perola model \nocite{jaffe73}(JP; {Jaffe} \& {Perola} 1973) assumes rapid isotropization
of the pitch angle distribution leading to an exponential cut-off in the CRE
spectrum. Further, energy dependent diffusion losses can also steepen the 
spectrum by 0.25, i.e., $\Delta \alpha = -0.25$ \nocite{condo92}({Condon} 1992). However, 
such losses would affect the CREs having high energies ($\gtrsim10$ GeV).

On the other hand, at low radio frequencies, the ionization losses have the
effect of flattening $\ant$. Under steady continuous injection of
electrons, the equilibrium spectra is flatter by 0.5, i.e., $\Delta \ant=+0.5$.
For single injection events, the resulting spectrum will depend on the ratio of
timescales of injection and ionization loss. Bremsstrahlung losses mildly
affects the CRE spectrum and therefore does not alter $\ant$ significantly
($\Delta \ant \approx 0$) \nocite{longa11}(see {Longair} 2011), both in the case of continuous
and single injection. Thus comparing the value of the observed $\ant$ with
respect to the typical injection values of $\sim-0.5$ to $-0.7$ one can broadly
get an idea of the dominant CRE energy loss mechanism.  Further, due to thermal
free--free absorption, the synchrotron spectrum is expected to turn over at
even lower frequencies below $\sim 0.1$ GHz.  These have a combined effect of
modifying the initial power-law spectrum into a curved spectrum with flatter
slope at low frequencies ($\lesssim1$ GHz) and steeper slope at high
frequencies ($\gtrsim1$ GHz).

The radio continuum emission from galaxies broadly originates due to two main
emission mechanisms -- 1) synchrotron emission  and 2) thermal
bremsstrahlung emission. Synchrotron emission contributes more than $\sim90$
percent to the total radio emission at low frequencies ($\lesssim1$ GHz)
\nocite{nikla97a, basu12a}({Niklas}, {Klein} \&  {Wielebinski} 1997; {Basu} {et~al.} 2012).  Contribution from thermal emission to the total
radio emission is significant at higher frequencies. For example, at 10 GHz,
$\sim40$ percent of the total emission is thermal in origin \nocite{gioia82}({Gioia}, {Gregorini} \& {Klein} 1982).
The thermal emission is also characterized by a power-law spectrum having
thermal spectral index $\alpha_{\rm th}=-0.1$ and is significantly flatter than
that of the $\ant$.  The composition of these two types of emission gives rise
to an overall power-law spectrum ($\alpha$) which is flatter than $\ant$. The
presence of thermal emission makes it difficult to interpret energy spectra of
the CREs and therefore it is necessary to subtract its contribution from the
total emission. Note that, in this paper we distinguish $\alpha$ and
$\ant$ as the spectral index of the total radio and synchrotron emission,
respectively.

A host of radio continuum surveys has been undertaken in the past to study
the broadband integrated spectra of galaxies and understand the various energy
loss mechanisms. Surprisingly, most of the studies showed a power-law spectrum
over broad range of frequencies. The power-law slope is found to be similar,
having a value $\sim-0.8$ with narrow dispersion, typically less than $\sim20$
percent. The power-law index, $\alpha$ between 1.4 and 5 GHz was found to be
$-0.85$ \nocite{srame75}({Sramek} 1975); between 0.4 and 10.7 GHz, $\alpha$ was found to be
$-0.71$ \nocite{klein81}({Klein} \& {Emerson} 1981) and $-0.74$ \nocite{gioia82}({Gioia} {et~al.} 1982).  Based on broadband
fitting of the radio continuum spectrum between 0.6 to 10 GHz for 41 spiral
galaxies, \nocite{duric88ea}{Duric}, {Bourneuf} \& {Gregory} (1988) found $\ant$ to peak at $\sim-0.8$. In a similar
study of 74 Shapely-Ames galaxies between 0.4 and 10.7 GHz, \nocite{nikla97a}{Niklas} {et~al.} (1997)
found the mean value of $\ant$ to be $-0.83$. In fact, for a much wider range
of radio frequencies, between 26.3 MHz and 22.8 GHz, \nocite{mulca14}{Mulcahy} {et~al.} (2014) did not
find any evidence of curvature in the radio continuum spectrum for the galaxy
M51. The spectrum is consistent with a single power-law with $\alpha =
-0.79\pm0.02$.

Recently, \nocite{marvi14}{Marvil}, {Owen} \& {Eilek} (2014), in a statistical study of 250 galaxies, found
evidence of curved radio continuum spectra between 74 MHz and 4.85 GHz.
However, there is no strong evidence of spectral curvature when studied for
integrated flux densities of individual galaxies except for a few cases, such
as M82 \nocite{condo92, adeba13}({Condon} 1992; {Adebahr} {et~al.} 2013), NGC 4631 \nocite{pohl90}({Pohl} \& {Schlickeiser} 1990), NGC 3627 and NGC 7331
\nocite{palad09}({Paladino}, {Murgia} \&  {Orr{\'u}} 2009), etc.  At much lower frequencies of 57.5 MHz, \nocite{israe90}{Israel} \& {Mahoney} (1990)
found evidence of systematically lower flux densities when compared to
extrapolated flux densities from higher frequencies assuming a single
power-law. They concluded that smoothly distributed diffuse ionized gas
throughout the galaxies is unlikely to cause thermal free--free absorption and
requires more clumpy medium of well-mixed non-thermal emitting and thermally
absorbing gas. 

\begin{table*} \centering 
 \caption{The sample galaxies.} 
  \begin{tabular}{@{}lcccccccc@{}} 
 \hline 
    Name  & Type  &Angular     & $i$& Distance   &  CO & H{\sc i} &Radio\\ 
          &       & size (D$_{25}$)($^\prime$) & ($^\circ$) & (Mpc) & && (1.4 GHz)\\ 
     (1)     &(2)       & (3) & (4) & (5)               & (6) & (7) & (8)\\ 
\hline 
NGC 4736    & SAab     & 11.2$\times$9.1  & 41 & 4.66$^1$     &HERACLES&THINGS     & Westerbork$^{b}$ SINGS$^4$\\ 
NGC 5055    & SAbc     & 12.6$\times$7.2  & 59 & 9.2$^\dagger$&HERACLES&THINGS& Westerbork SINGS$^4$\\ 
NGC 5236    & SABc     & 11.2$\times$11 & 24 & 4.51$^2$ &NRAO 12 m&THINGS&  VLA$^a$ CD array$^5$\\ 
NGC 6946    & SABcd    & 11.5$\times$9.8  & 33 & 6.8$^3$       &HERACLES&THINGS&  VLA C+D array$^6$\\ 
\hline 
\end{tabular}

Column 3 lists the optical diameter measured at the 25 magnitude arcsec$^{-2}$
contour.  The inclination angles ($i$; $0^\circ$ is face-on) are listed in
Column 4.  Distances in column 5 are taken from: $^1$ \nocite{karac03}{Karachentsev} {et~al.} (2003), $^2$
\nocite{karac02}{Karachentsev} {et~al.} (2002), $^3$ \nocite{karac00}{Karachentsev}, {Sharina} \&  {Huchtmeier} (2000) and the NED $^\dagger$. Columns 6 and 7
list the data used to trace the molecular and atomic gas, respectively. Column
8 lists the sources of archival data at 1.4 GHz: $^4$ \nocite{braun07}{Braun} {et~al.} (2007), $^5$ VLA
archival data using CD array (project code: AS325), $^6$ VLA map obtained by
combining archival data from C and D array \nocite{beck07}({Beck} 2007).\\ $^a$ The Very
Large Array (VLA) is operated by the NRAO. The NRAO is a facility of the
National Science Foundation operated under cooperative agreement by Associated
Universities, Inc.\\ $^b$ The Westerbork Synthesis Radio Telescope (WSRT) is
operated by the Netherlands Foundation for Research in Astronomy (NFRA) with
financial support from the Netherlands Organization for scientific research
(NWO). 
\label{sampletab} 
\end{table*}

Although the global spectral index, for both total and non-thermal emission, of
galaxies does not vary and is consistent with $\ant\sim-0.8$, locally they show
large variations.  The spectral index is observed to vary significantly
radially, being flatter and close to the injection value ($\ant \sim -0.5$
to $-0.6$) towards the central regions  and steepens towards the outer parts
($\ant\sim -0.9$ to $-1.2$) \nocite{basu12a, beck07, tabat07, palad09}({Basu} {et~al.} 2012; {Beck} 2007; {Tabatabaei} {et~al.} 2007; {Paladino} {et~al.} 2009).  The
various CR energy loss mechanisms depend on local physical parameters of the
ISM such as the magnetic field strength, the number density of neutral gas,
energy density of the photon field, shocked regions, etc.  Moreover, the
magnetic field strengths and gas density are believed to be coupled due to
magnetohydrodynamic turbulence in the ISM \nocite{chand53, cho00, grove03}({Chandrasekhar} \& {Fermi} 1953; {Cho} \& {Vishniac} 2000; {Groves} {et~al.} 2003).
Further, injection rate of CR particles depends on the star formation rate
which is related to the gas density through the Kennicutt-Schmidt law
\nocite{kenni98}({Kennicutt} 1998).  Local ISM densities can therefore, directly or indirectly
shape up the spectral index differently giving rise to large variations.  It is
therefore imperative to study the spatially resolved non-thermal spectral index
and compare it to the local ISM gas densities. In this paper we qualitatively
discuss the effects of gas surface densities on the non-thermal spectral index
and the most likely scenario of CR energy loss prevalent in spatially resolved
galaxies.

The paper is organized as follows: In Section 2 we present the various source
of the data used. The galaxies studied are NGC 4736, NGC 5055, NGC 5236 and NGC
6946. We present our results in Section 3 and discuss them in Section 4. 
The conclusions based on this work are summarized in Section 5.

\begin{table*}
\caption{Resolution of the available maps in arcsec$^2$.}
\begin{centering}
\begin{tabular}{@{}lcccccc@{}}
\hline
Name & 0.33-GHz & 1.4-GHz  & H{\sc i}  & CO  & $\ant$ & Linear\\
     &          &	   &	       &     & maps   & scale (kpc) \\
\hline
NGC 4736 & 13$\times$12	& 19$\times$12.5   & 10.2$\times$9.1  & 13.4$\times$13.4 & 20$\times$20 & 0.45\\
NGC 5055 & 17$\times$10	& 18.5$\times$12.5 & 10.1$\times$8.7  & 13.4$\times$13.4 & 20$\times$20 & 0.90\\
NGC 5236 & 16$\times$12	& 26$\times$14	   & 15.2$\times$11.4 & 55$\times$55 & 26$\times$14     & 0.55\\
NGC 6946 & 12$\times$11	& 15$\times$15     & 6$\times$5.6     & 13.4$\times$13.4 & 15$\times$15 & 0.50\\
\hline
\end{tabular}
\end{centering}
\label{resolution}
\end{table*}

\section{Data and analysis}

In this work we study four nearby galaxies, namely, NGC 4736, NGC 5055, NGC
5236 and NGC 6946.  The galaxies are chosen from \nocite{basu12a}{Basu} {et~al.} (2012) 
and were observed using the Giant Metrewave Radio Telescope (GMRT) at 0.33
GHz.  To measure the spectral index, we used archival data at a higher radio
frequency near 1.4 GHz.  Table~\ref{sampletab} lists the salient features of
the sample and the sources of the archival data. As pointed out earlier, the
radio continuum emission originates from both synchrotron emission and thermal
free--free emission.  Owing to the steep spectral index, $\ant<-0.5$, the
synchrotron emission is expected to dominate over the thermal emission,
especially at lower ($<1$ GHz) radio frequencies. However, due to high
star-formation activity and/or certain bright \hii regions in galaxies, the
contribution of the thermal emission can be significant locally
\nocite{basu12a}({Basu} {et~al.} 2012). The contamination due to the thermal component of the radio
emission, which has significantly flatter spectral index of $-0.1$, affects the
synchrotron spectral index, making it flatter than actual by more than 30
percent \nocite{basu12a}(see Figure 6 of {Basu} {et~al.} 2012). This necessitates a thorough
separation of the thermal emission from the total radio emission for robust
determination of $\ant$.

We used the recent method of estimating the thermal component of the radio
emission given in \nocite{basu13}{Basu} \& {Roy} (2013). The thermal emission is determined by
combining H$\alpha$ and $\lambda24~\mu$m emission from galaxies as its tracer.
This method allows us to probe the non-thermal emission from galaxies at linear
scales of 0.4--1 kpc, depending on the distance of the galaxies.
Table~\ref{resolution} lists the angular resolution of the available $\ant$
maps. The overall thermal fraction\footnote{Thermal fraction ($f_{\rm th,\nu}$)
at a radio frequency $\nu$ is defined as: $f_{\rm th,\nu} = S_{\rm
th,\nu}/S_{\rm tot, \nu}$. Here, $S_{\rm th,\nu}$ and $S_{\rm tot, \nu}$ are
the estimated thermal emission and the total radio emission, respectively.} at
0.33 GHz ($f_{\rm th, 0.33GHz}$) was found to be less than 5 percent while at
1.4 GHz ($f_{\rm th, 1.4GHz}$) it is found to be $\sim10$ percent
\nocite{basu13, basu12a}({Basu} \& {Roy} 2013; {Basu} {et~al.} 2012). However, locally, $f_{\rm th, 0.33GHz}$ and $f_{\rm
th, 1.4GHz}$ are as high as 15 percent and 30 percent, respectively.

\begin{figure*}
	\begin{tabular}{cc}
	{\mbox{\includegraphics[width=7cm]{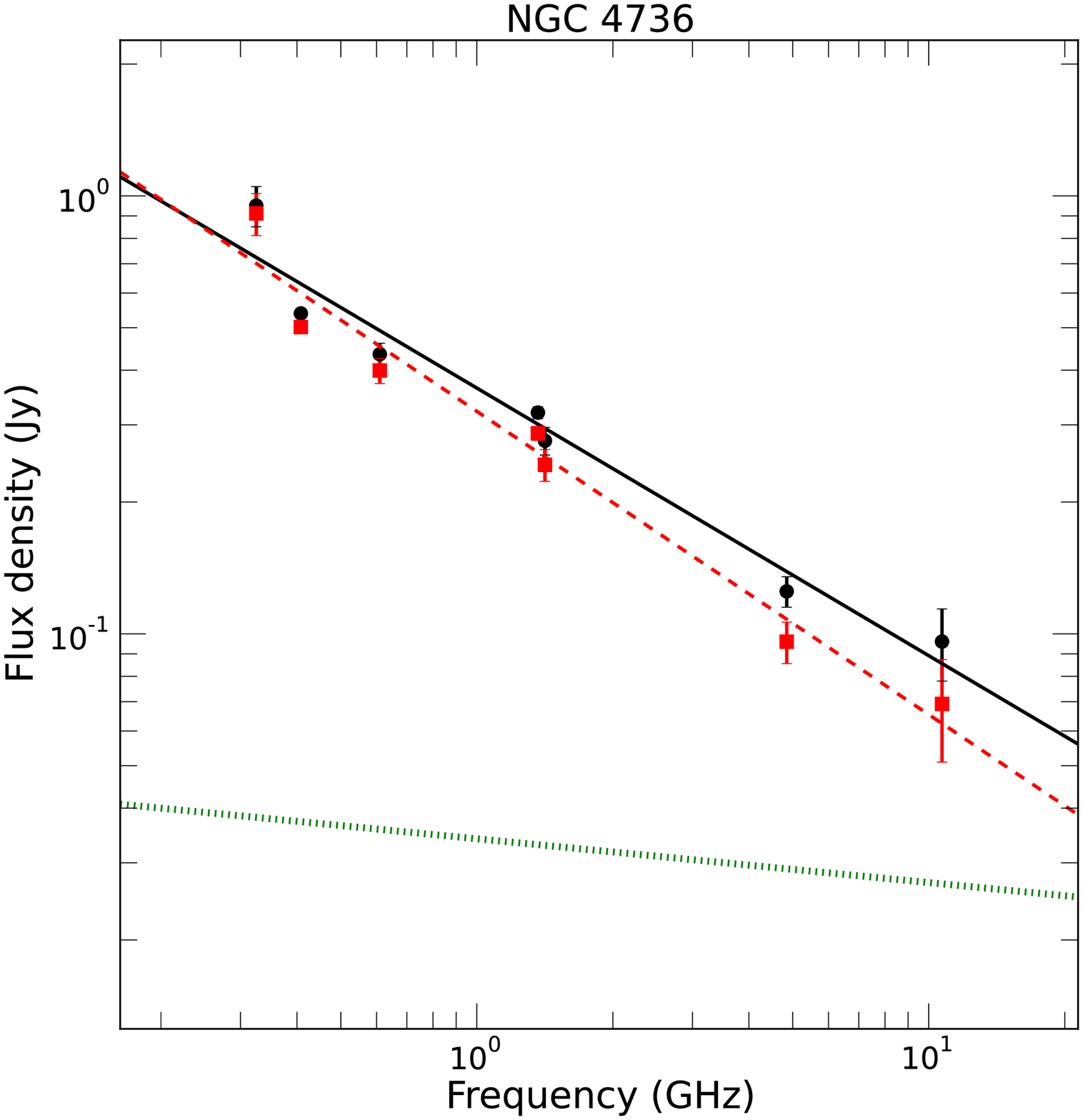}}} &
	{\mbox{\includegraphics[width=7cm]{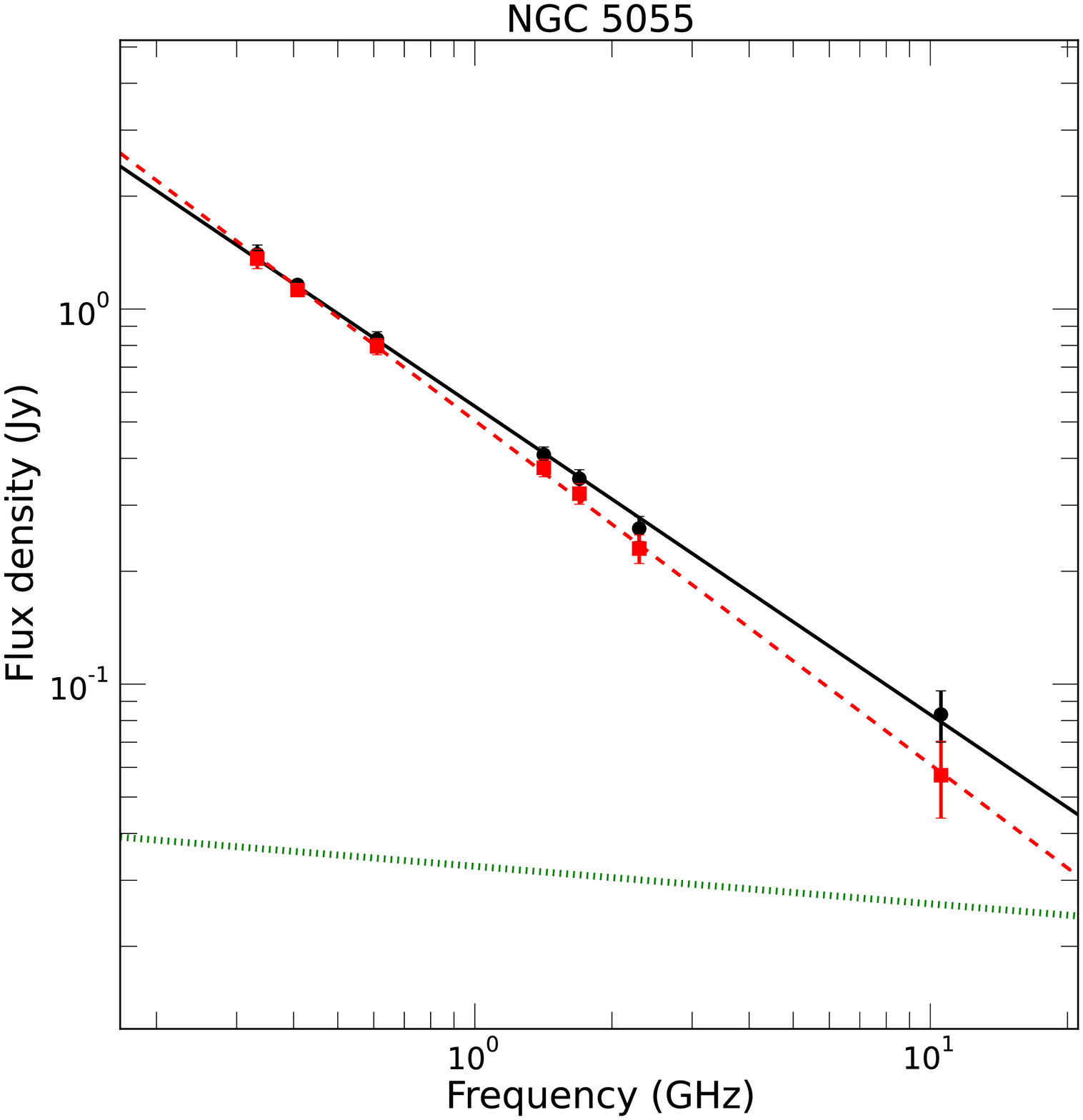}}} \\
	{\mbox{\includegraphics[width=7cm]{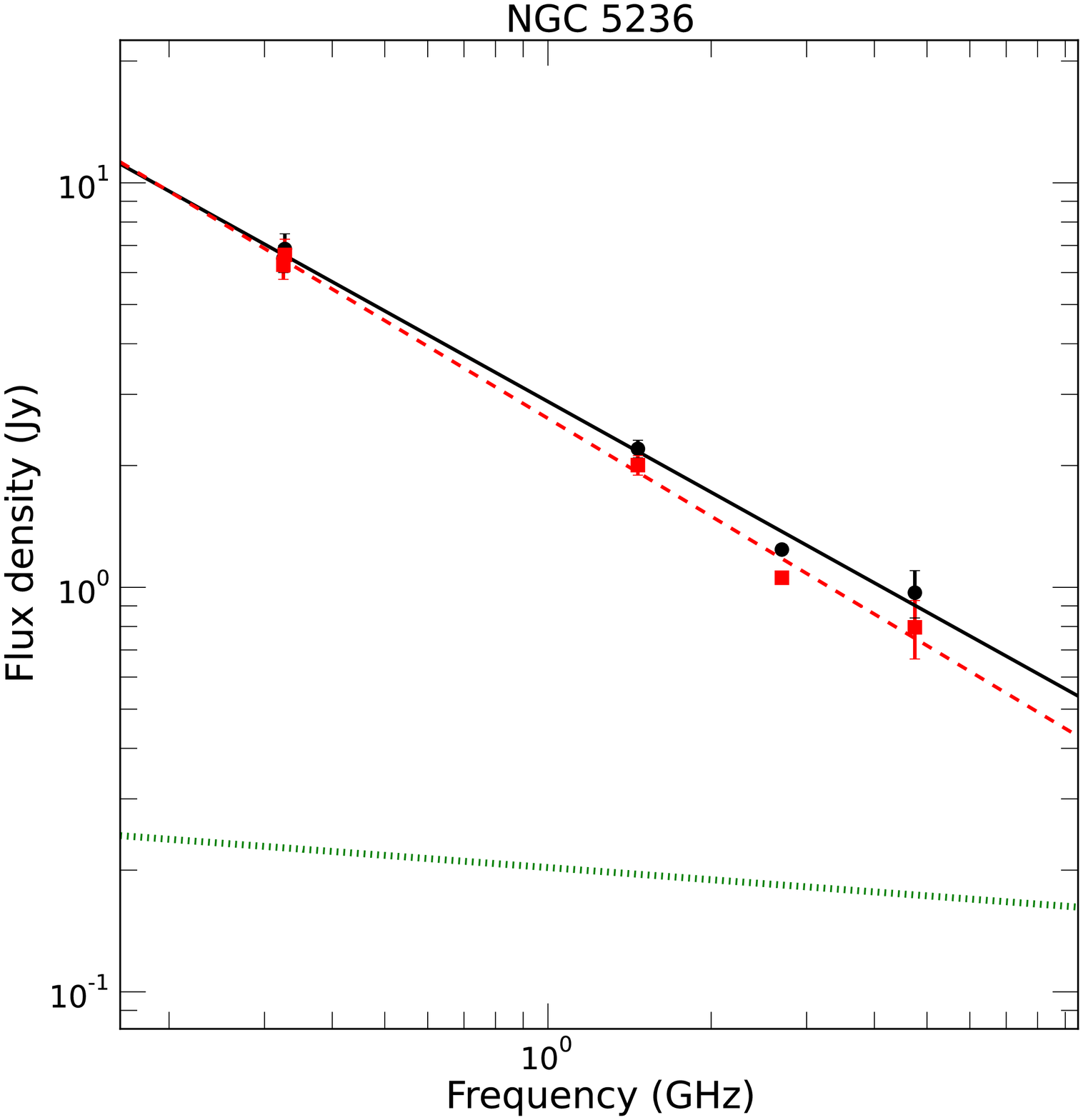}}} &
	{\mbox{\includegraphics[width=7cm]{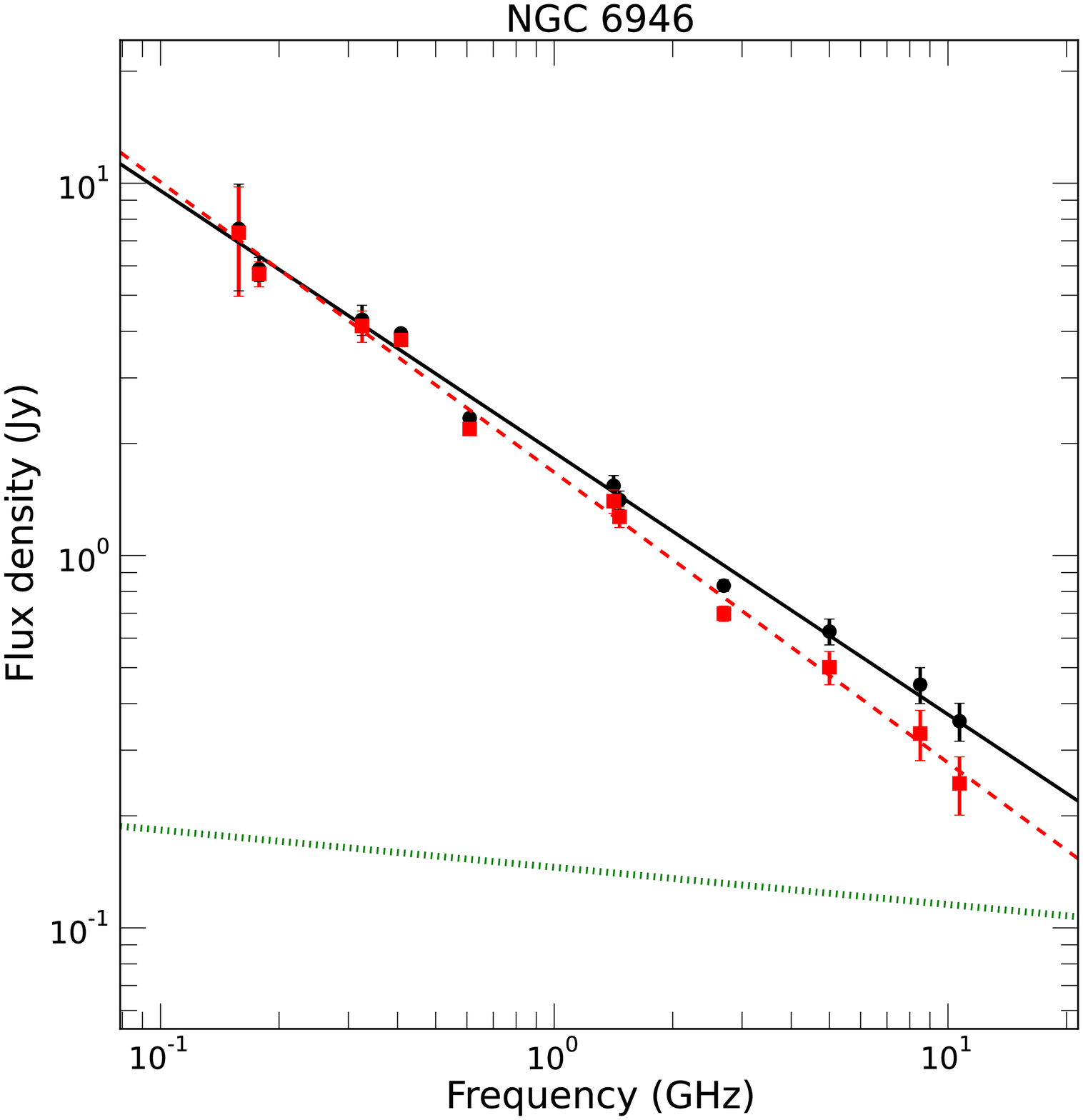}}} \\
	\end{tabular}
\caption{Integrated radio continuum spectrum of the sample galaxies. The black
dots show the total observed flux density and the red squares show the
estimated non-thermal flux density. The solid and dashed lines show the
power-law fit to the total and non-thermal emission respectively. The green
dotted lines show the extrapolated thermal emission.}
\label{integ}
\end{figure*}

To estimate the surface mass density of total neutral gas ($\Sigma_{\rm gas}$) in
the galaxies, we used H{\sc i} and CO line emission as tracers of the neutral
atomic and molecular gas components, respectively.  We used the
natural-weighted H{\sc i} moment-0 maps from the H{\sc i} Nearby Galaxy Survey
\nocite{walte08}(THINGS; {Walter} {et~al.} 2008) to trace the atomic gas for all our sample galaxies.
The galaxies were observed with high angular resolution of $\sim10$ arcsec (see
Table~\ref{resolution}) and velocity resolution $\leq5.2$  km s$^{-1}$ using
the VLA.  We used CO(J2$\to$1) emission from the galaxies, NGC 4736, NGC 5055
and NGC 6946 as a tracer of the molecular gas in the ISM. We used the
CO(J2$\to$1) moment-0 maps from the HERA CO Line Extragalactic Survey
\nocite{leroy09}(HERACLES; {Leroy} {et~al.} 2009). The galaxies were observed with an angular
resolution of $\sim13$ arcsec and velocity resolution of 2.6 km s$^{-1}$ using
the IRAM 30-m telescope. For NGC 5236, we used CO(J1$\to$0) line emission as
the tracer of molecular gas observed with the NRAO 12-m telescope
\nocite{crost02}({Crosthwaite} {et~al.} 2002). The CO(J1$\to$0) moment-0 map has an angular resolution of 55
arcsec and velocity resolution of 5.2 km s$^{-1}$.

We computed total neutral gas surface mass density maps on a pixel-by-pixel
basis using the surface mass density maps of atomic ($\Sigma_{\rm H{\sc I}}$)
and molecular gas ($\Sigma_{\rm H_2}$) as, $\Sigma_{\rm gas} = 1.36 \times
(\Sigma_{\rm H{\sc I}} + \Sigma_{\rm H_2})$. The 1.36 factor is to account for
the presence of Helium. The H{\sc i} gas surface mass density was estimated
using,
\begin{equation}
\Sigma_{\rm H{\sc I}} ({\rm M_\odot pc^{-2}}) = 0.015~\cos i ~ I_{\rm H{\sc I}} ~ ({\rm K~km~s^{-1}}).
\end{equation}
Here, $i$ is the inclination of the galaxy, defined such that $i=0$ is face-on
and $I_{\rm H{\rm I}}$ is the line-integrated intensity.

Similarly, the CO(J2$\to$1) line-integrated intensity, $I_{\rm CO_{2\to 1}}$,
of the HERACLES maps were converted to the molecular gas surface mass density
using, \begin{equation} \Sigma_{\rm H_2} ({\rm M_\odot pc^{-2}}) = 5.5~\cos i ~
I_{\rm CO_{2\to 1}} ~ ({\rm K~km~s^{-1}}).  \end{equation} This assumes a
constant CO-to-H$_2$ conversion factor, $X_{\rm CO} = 2\times 10^{20} {\rm
cm^{-2}~({\rm K~km~s^{-1}})^{-1}}$ for the CO(J1$\to$0) transition and a line
ratio, $I_{\rm CO_{2\to 1}}/I_{\rm CO_{1\to 0}}=0.8$. The line-integrated flux
density, $S_{\rm CO_{1\to 0}}$, of the CO(J1$\to$0) moment-0 NRAO 12-m map for
the galaxy NGC 5236 was converted to molecular gas mass, $M_{\rm H_2}$
\nocite{young89}({Young} {et~al.} 1989), 
\begin{equation} M_{H_2} ({\rm M_\odot}) = 1.1\times 10^4
	D_{\rm Mpc}^2~\cos i~S_{\rm CO_{1\to 0}} ({\rm Jy~km~s^{-1}}).
\end{equation} 
Here, $ D_{\rm Mpc}$ is the distance to the galaxy in Mpc. The molecular gas
mass was converted to $\Sigma_{\rm H_2}$ by $\Sigma_{\rm H_2} ({\rm M_\odot
pc^{-2}}) = M_{\rm H_2} ({\rm M_\odot})/[x_{\rm pix}\times y_{\rm pix}~(\rm
pc^{2})]$.  Here, $x_{\rm pix}$ and $y_{\rm pix}$ are the pixel sizes of the
map expressed in pc at the distance of the galaxy.

\begin{figure*} 
\begin{tabular}{cc}
{\mbox{\includegraphics[width=7cm]{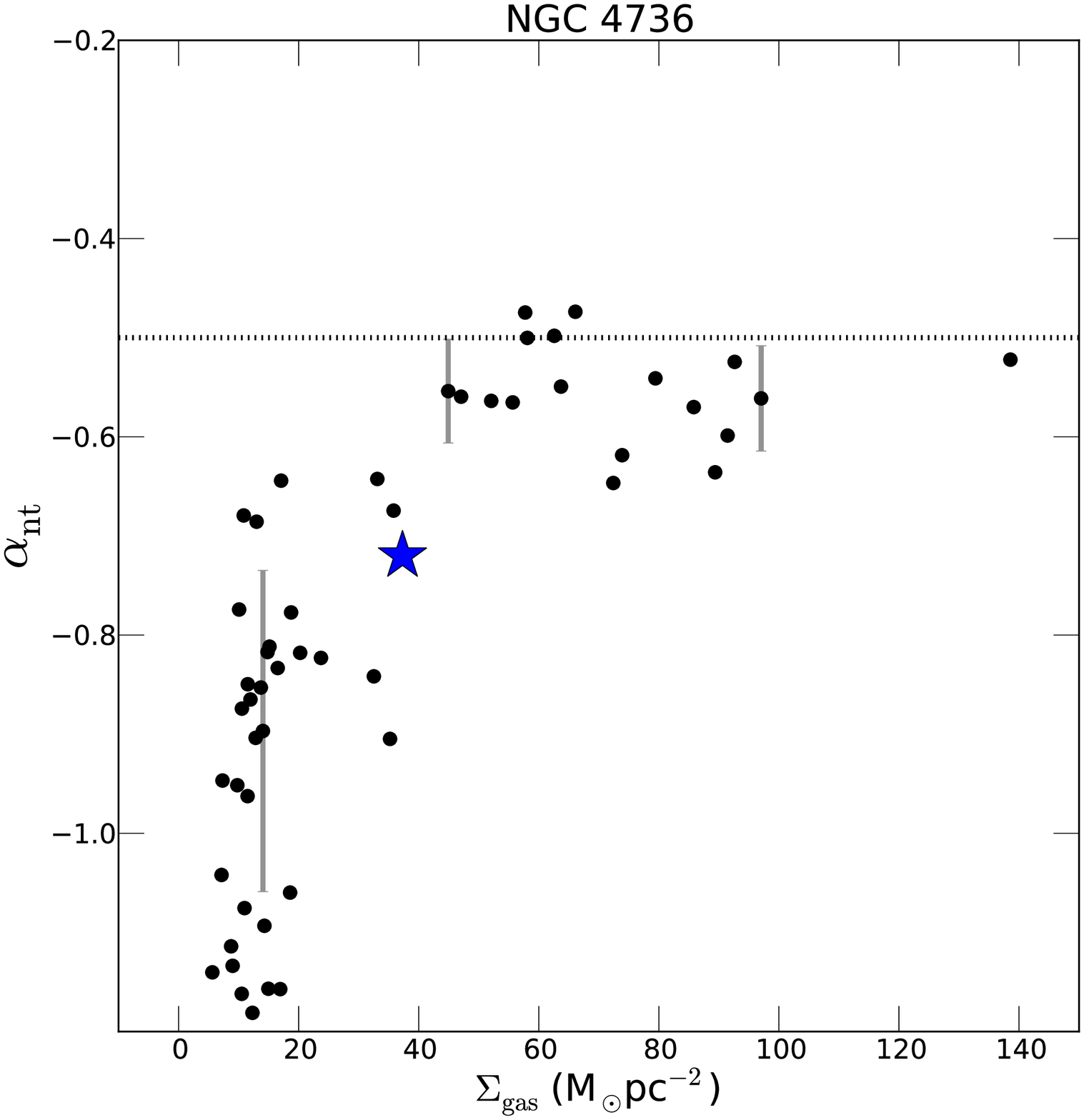}}} &
{\mbox{\includegraphics[width=7cm]{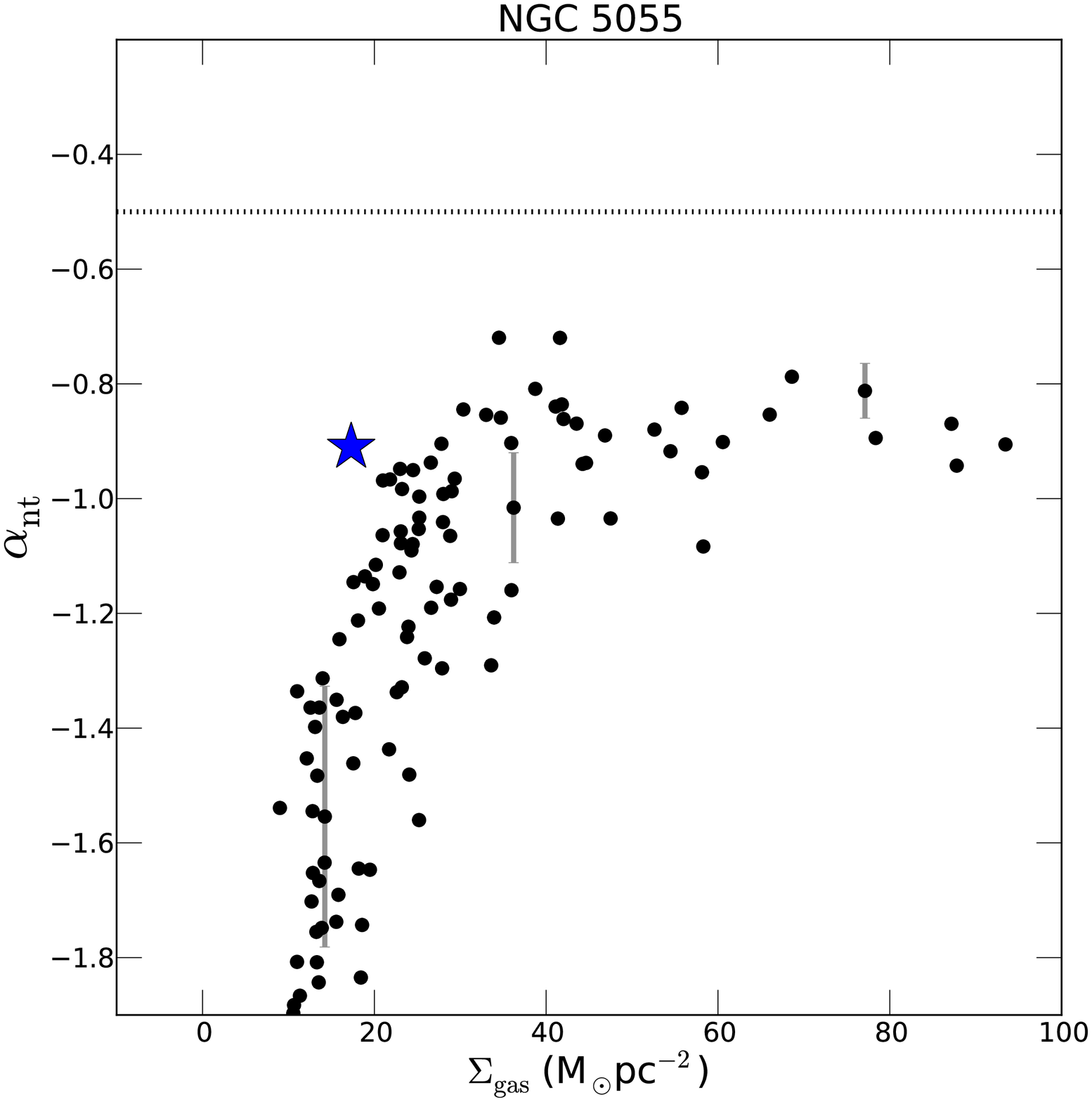}}} \\
{\mbox{\includegraphics[width=7cm]{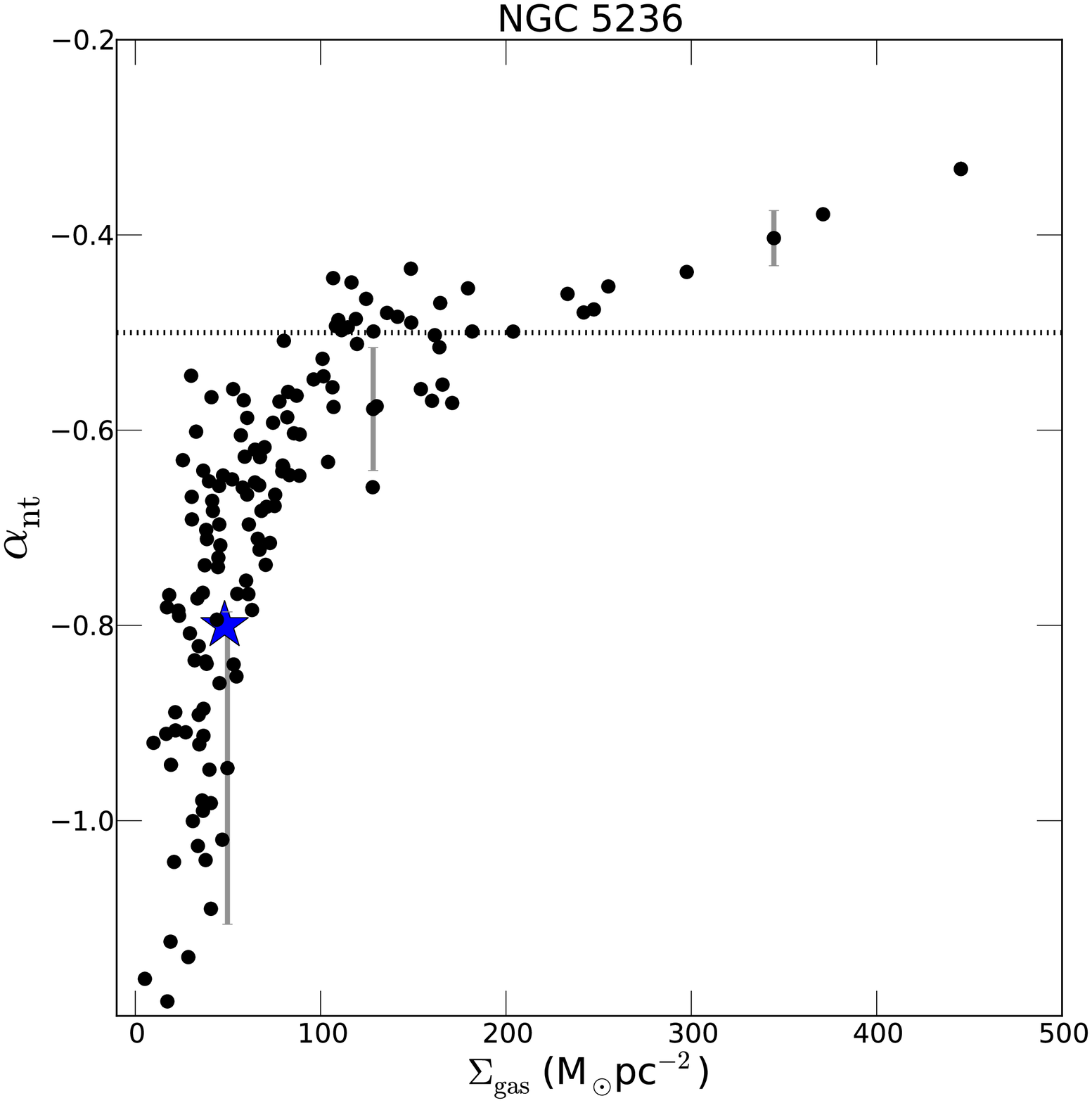}}} &
{\mbox{\includegraphics[width=7cm]{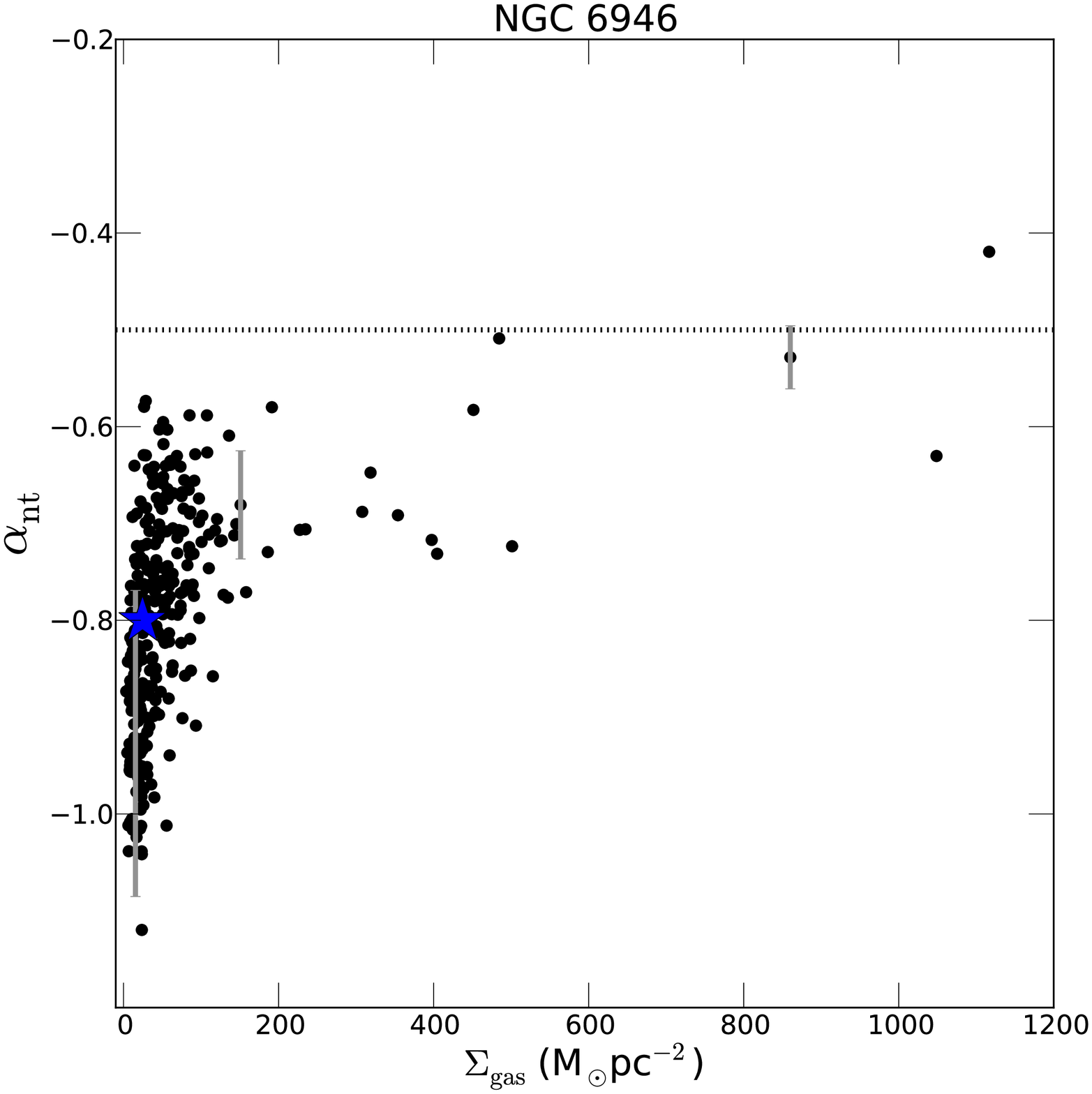}}} \\ 
\end{tabular}
\caption{Non-thermal spectral index ($\ant$) versus the total neutral gas
(atomic + molecular) surface density ($\Sigma_{\rm gas}$) in units of $\rm
M_\odot pc^{-2}$. The dotted lines shows $\ant=-0.5$. The blue stars shows the
galaxy averaged values. We show representative error bars for some of the
points covering the range of $\Sigma_{\rm gas}$ for each galaxy.}
\label{ant-gas}
\end{figure*}

\section{Results}

Figure~\ref{integ} shows the galaxy integrated radio continuum spectrum for the
four galaxies between 0.33 and 10.7 GHz. The black dots represent the total
observed flux densities obtained from various sources in the literature
\nocite{basu12a}(see Table~3 of {Basu} {et~al.} 2012). The higher frequency ($\gtrsim 5$ GHz) data
points were mostly measured using single dish observations and hence, do not
suffer from missing flux issues.  The broad-band spectrum was fitted using 2
types of models in $\log-\log$ space: 1) a single power-law and 2) a second
order polynomial that could account for any curvature in the spectrum, if
present.  The single power-law has the form, $\log S_\nu = c + \alpha \log
\nu$.  The second order polynomial has the form, $\log S_\nu = c + \alpha \log
\nu + \beta (\log \nu)^2$.  Here, $c$, $\alpha$ and $\beta$ are the fitting
parameters.  

For all the galaxies, the best fit model is well represented by a single
power-law shown as the solid black lines in Figure~\ref{integ}. The power-law
index, $\alpha$, is found to be $-0.62\pm0.05$, $-0.82\pm0.01$, $-0.74\pm0.02$
and $-0.70\pm0.02$ for the galaxies NGC 4736, NGC 5055, NGC 5236 and NGC 6946,
respectively. The red squares show the non-thermal flux densities estimated by
subtracting the thermal emission (shown as the green dotted lines) at the
respective frequencies.  The non-thermal emission spectrum for all the four
galaxies can also be well fitted by a power-law having index steeper than the
total emission. The non-thermal power-law index, $\ant$, is estimated to be
$-0.70\pm0.05$, $-0.91\pm0.01$, $-0.80\pm0.03$, $-0.82\pm0.02$ for the galaxies
NGC 4736, NGC 5055, NGC 5236 and NGC 6946, respectively.  Within the
uncertainty of the measurements, we do not find any steepening of the overall
spectrum towards higher frequencies ($\gtrsim2$ GHz) as expected due to
synchrotron and/or inverse-Compton losses. Towards, the lower frequencies
($\lesssim0.5$ GHz), we do not observe any flattening due to ionization losses
and/or thermal free--free absorption. Our result is similar to what was observed
for the galaxy M51 over a much wider range of frequencies \nocite{mulca14}({Mulcahy} {et~al.} 2014).

In Figure~\ref{ant-gas} we plot the variation of the $\ant$ with total surface
density of the neutral (atomic+molecular) gas ($\Sigma_{\rm gas}$) for all the
four galaxies. $\ant$ and $\Sigma_{\rm gas}$ is computed within one synthesized
beam given in Table~\ref{resolution}.  To ensure independence, each point is
measured, roughly separated by one synthesized beam. The $\ant$ was computed
using the non-thermal maps between 0.33 and 1.4 GHz considering pixels having
signal-to-noise ratio more than 3$\sigma$ at both the frequencies. Each beam
corresponds to $0.5-1$ kpc in linear size depending on the distance of the
galaxies (see Table~\ref{resolution}). For the galaxy NGC 5236, due to poor
resolution of the available CO map, the linear resolution was limited to $1.2$
kpc.

We find that in the arm regions, $\ant$ is flatter than that in the interarm
regions. This was already indicated in \nocite{basu12a}{Basu} {et~al.} (2012) and \nocite{tabat07}{Tabatabaei} {et~al.} (2007).
There appears to be a sharp transition of spectral index above and below
$\sim50~\rm M_\odot pc^{-2}$. We find the spectral index to remain constant or
get flatter with increasing $\Sigma_{\rm gas}$ above $\sim50~\rm M_\odot
pc^{-2}$. There is a sharp steepening of the spectral index below $\sim50~\rm
M_\odot pc^{-2}$, typically in outer parts and interarm regions. We find
$\ant$ close to the injection value of $\sim-0.5$ for regions having
$\Sigma_{\rm gas}$ in the range $50-250~\rm M_\odot pc^{-2}$.  At even denser
regions, we see the $\ant$ to be flatter than $-0.5$ for the galaxy NGC 5236.

The error on $\ant$ depends on the signal-to-noise ratio (S/N) of the flux
densities within each synthesized beam at each radio frequency. The regions of
high gas density ($\Sigma_{\rm gas} \gtrsim 50~\rm M_\odot pc^{-2}$) are also
typically the regions with strong radio emission with S/N $\gtrsim5$.  Such
regions have $\lesssim5$ percent error in the $\ant$. While the low gas density
regions ($\Sigma_{\rm gas} \lesssim 50~\rm M_\odot pc^{-2}$) with radio flux
density S/N $\lesssim5$, $\ant$ have up to 20 percent error.  Additional error
on the estimated value of $\ant$ is incurred due to uncertainties in the
estimated thermal emission.  This is primarily due to the unknown value of the
electron temperate ($T_e$), assumed to be $10^4$ K, while estimating the
thermal emission \nocite{tabat07}(see {Tabatabaei} {et~al.} 2007). As a result, the uncertainty in the
thermal fraction ($f_{\rm th}$) is $\sim10$ and $\sim15$ percent at 0.33 and
1.4 GHz, respectively \nocite{tabat07, basu12a}({Tabatabaei} {et~al.} 2007; {Basu} {et~al.} 2012).  This introduces an error
of $\sim 5$ percent on the $\ant$.  However, note that the uncertainty in
$f_{\rm th}$ affects $\ant$ only towards the inner parts of the galaxies where 
$f_{\rm th}$ is high \nocite{basu12a}({Basu} {et~al.} 2012).  The $\ant$ in the outer parts are not
affected due to low thermal emission.  Overall, the estimated $\ant$ have
maximum error of $\sim7$ percent towards the inner parts of the galaxies, i.e.,
regions of higher $\Sigma_{\rm gas}$ and up to $\sim20$ percent in the outer
parts, i.e., regions of lower $\Sigma_{\rm gas}$. We show representative errors
on $\ant$ for some of the points covering the span of $\Sigma_{\rm gas}$ for
each galaxy in Figure~\ref{ant-gas}.

We checked the possibility of free--free absorption of the synchrotron emission
by thermal electrons giving rise to the observed flattening at lower
frequencies. The observed flux density can be expressed as $S_\nu \propto
\nu^{\ant} \exp(-\tau_{\rm ff})$.  $\tau_{\rm ff}$ is the free--free optical
depth at a radio frequency $\nu$ and is given by, 
\begin{equation}
	\tau_{\rm ff} = 0.082~T_e^{-1.35}\left(\frac{\nu}{\rm GHz}\right)^{-2.1} \left(\frac{EM}{\rm cm^{-6} pc}\right)
\end{equation}
Here, $T_e$ is the temperature of the thermal electrons assumed to be $10^4$ K
and $EM$ is the emission measure. The emission measure for our sample galaxies
are in the range of few 100s (in the disk and outer parts of the galaxies) to
10$^4~{\rm cm^{-6}pc}$ (towards the center and H{\sc ii} regions)
\nocite{basu12a}({Basu} {et~al.} 2012).  For this range of $EM$s, the $\tau_{\rm ff}$ is expected to
become 1, below 0.1 GHz.  At 0.33 GHz, $\tau_{\rm ff}$ for our sample galaxies
lies in the range $10^{-3}$ to few times $10^{-2}$.  Moreover, for free--free
absorption to affect at 0.33 GHz, $EM \approx 10^6~{\rm cm^{-6}pc}$ is required.
Regions with such extreme $EM$s are rare in galaxies except perhaps in the
galactic centers \nocite{roy04}({Roy} \& {Rao} 2004). The flattening of the radio continuum spectra
at 0.33 GHz is therefore unlikely to be caused by free--free absorption.

\section{Discussion}

\begin{figure*} 
\begin{center} 
\begin{tabular}{cc}
\includegraphics[width=7cm]{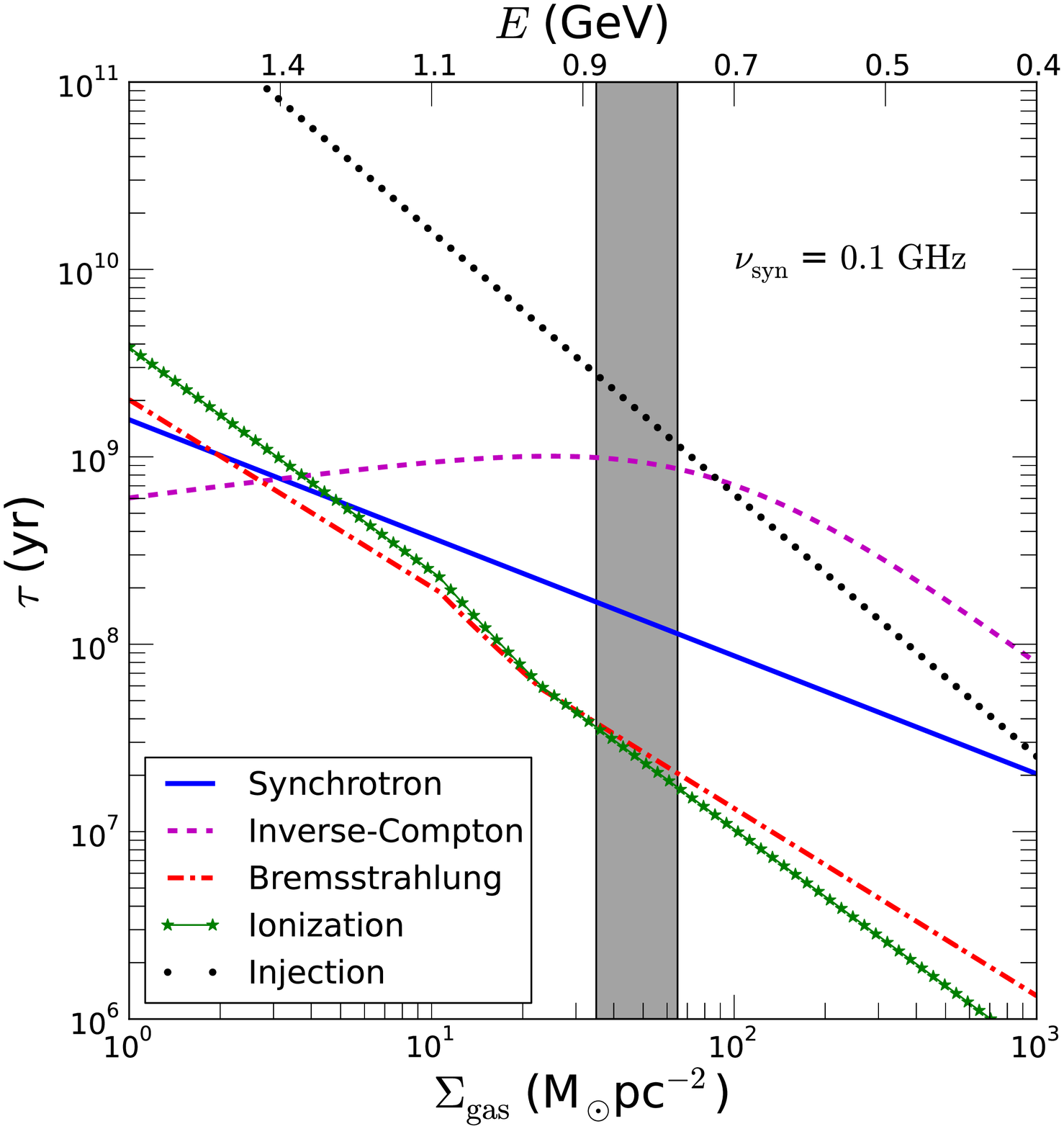} &
\includegraphics[width=7cm]{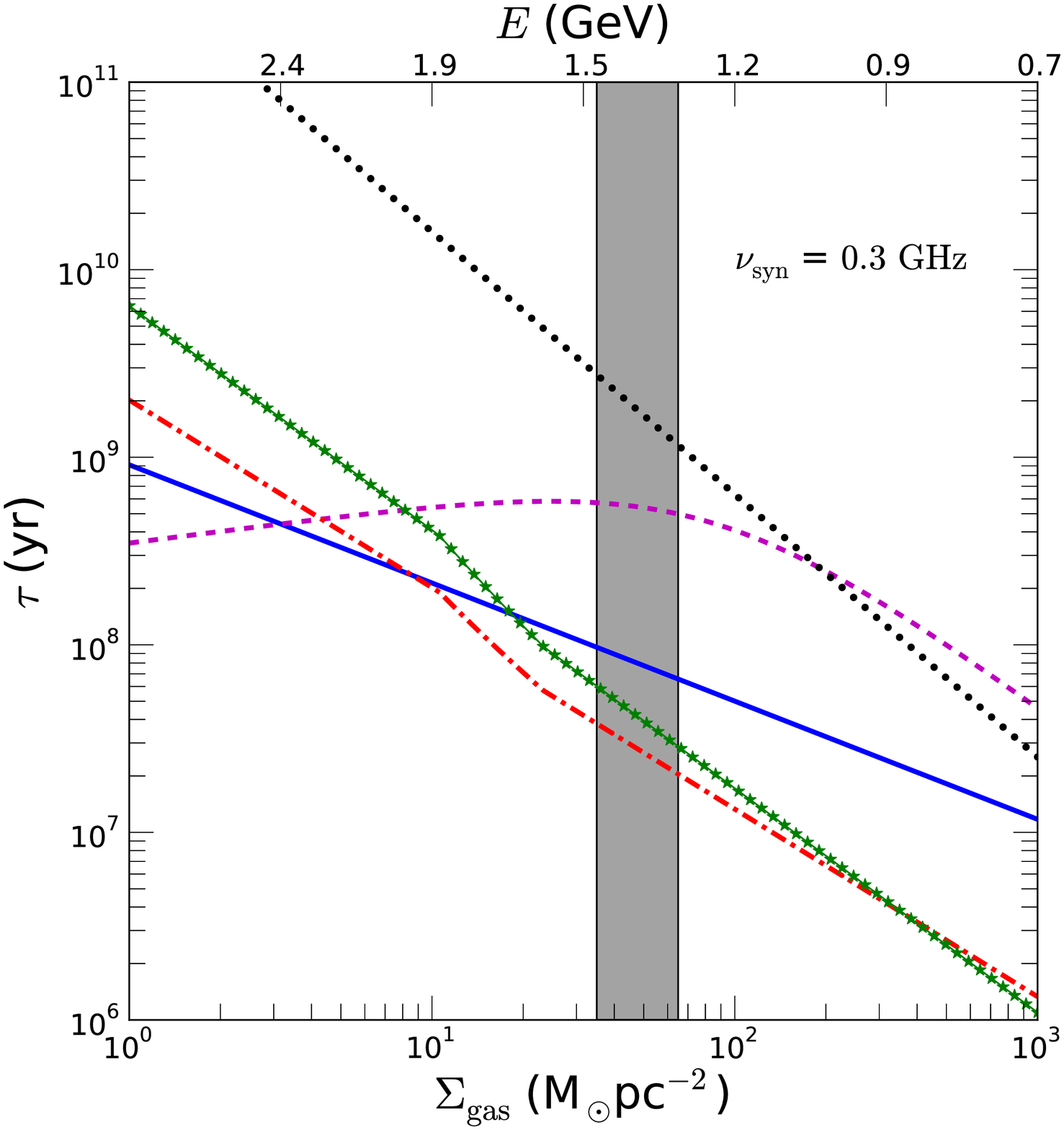} \\
\includegraphics[width=7cm]{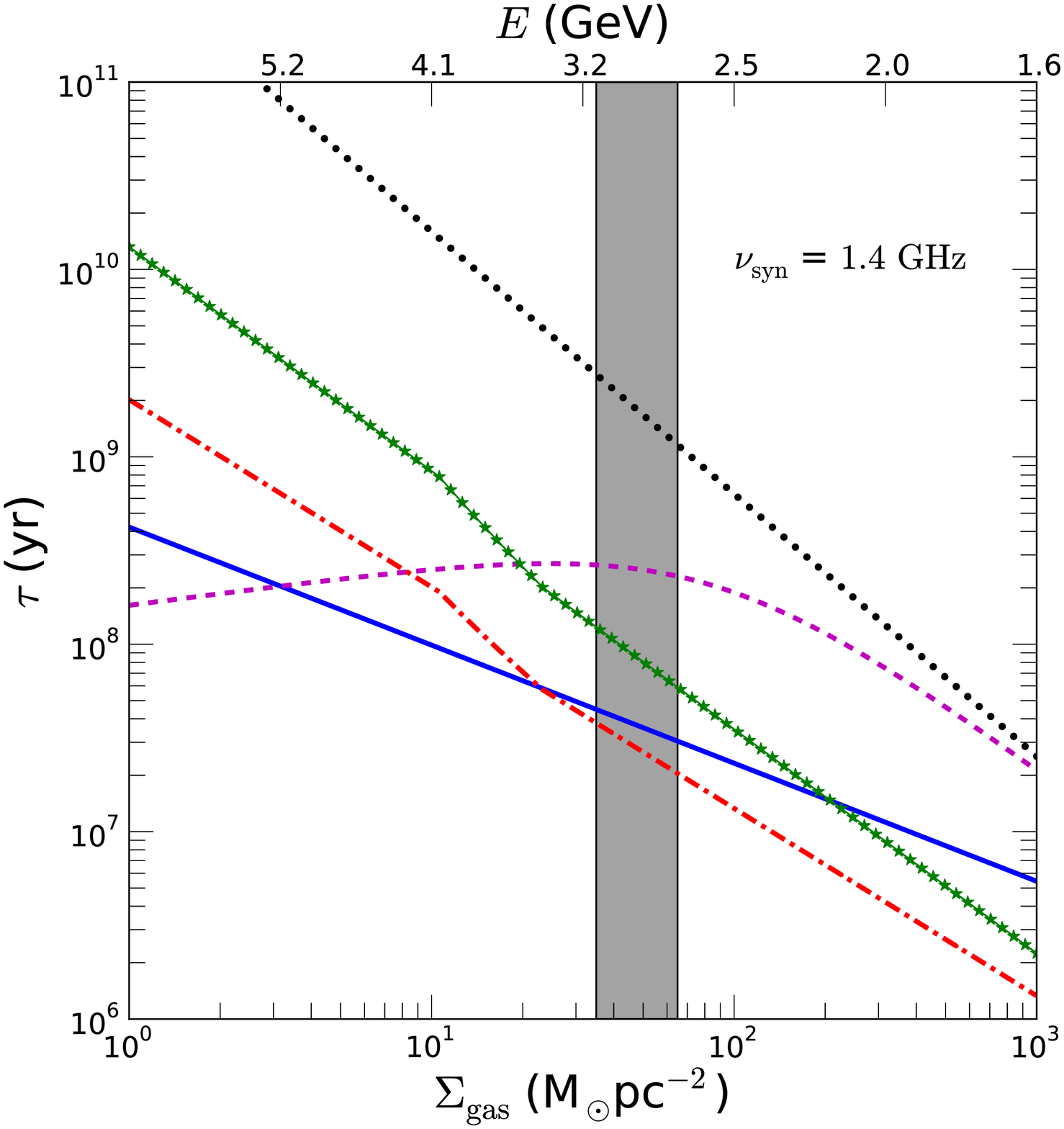} &
\includegraphics[width=7cm]{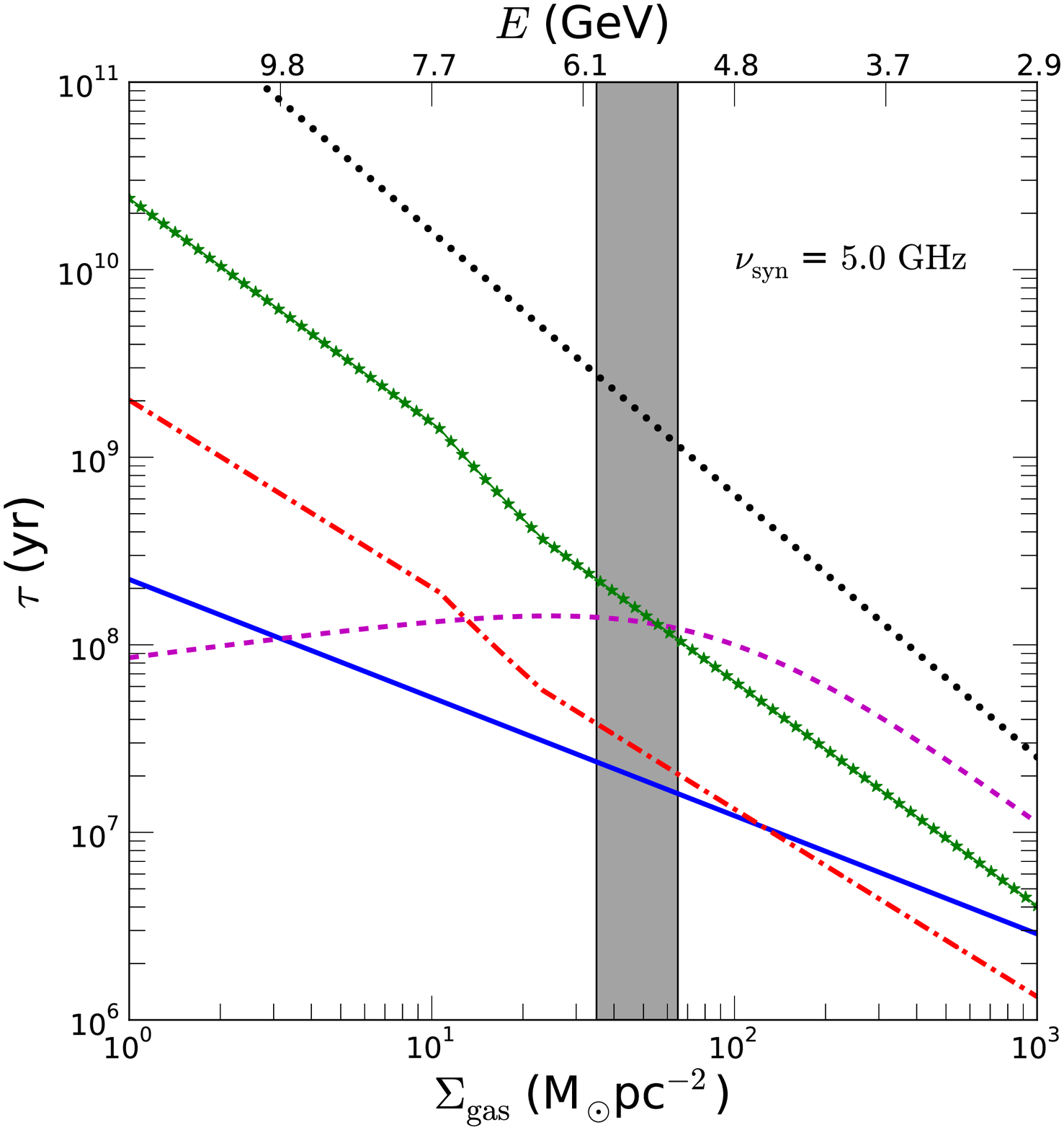}\\ 
\includegraphics[width=7cm]{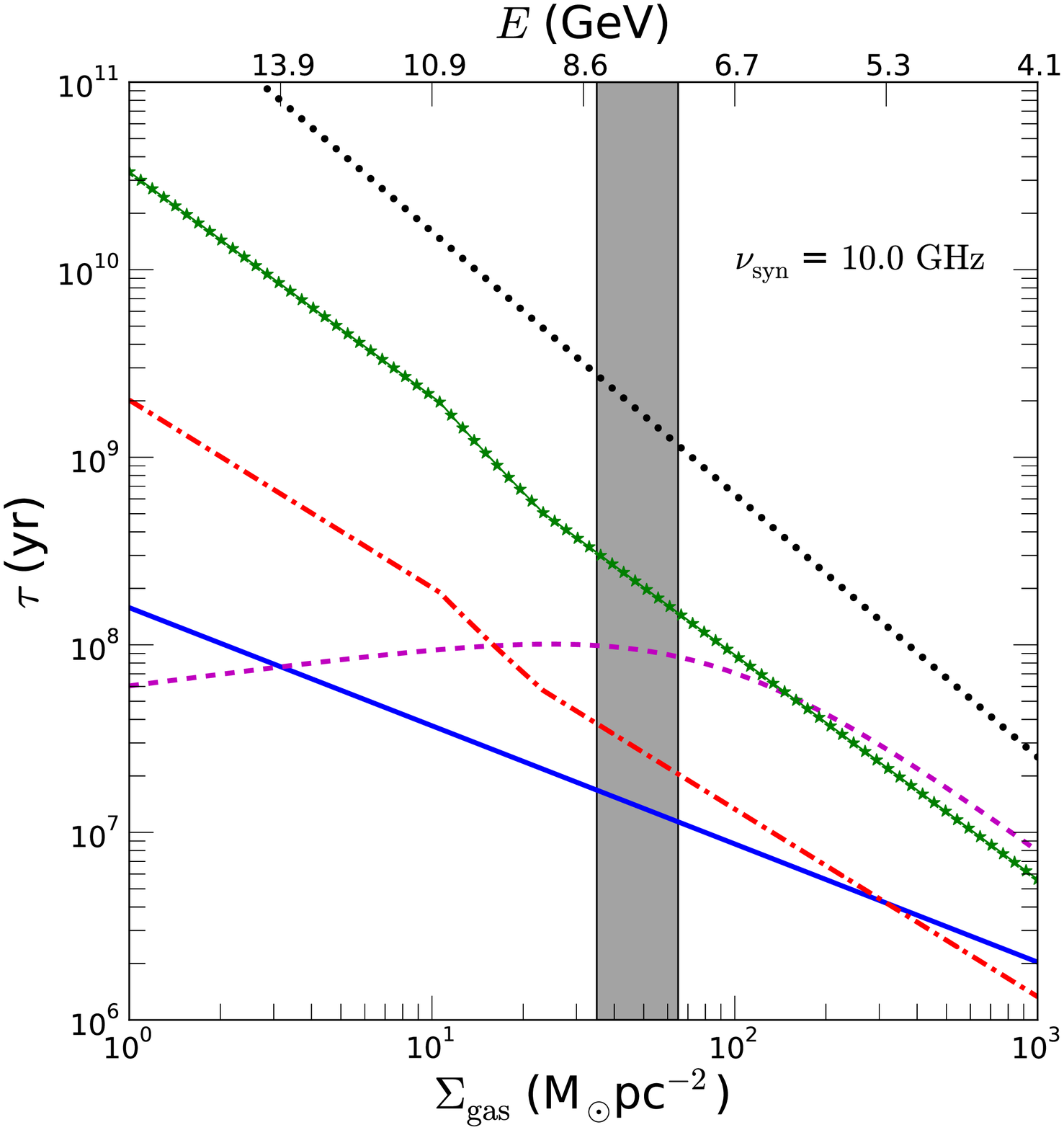}&
\\
\end{tabular} 
\end{center}
\caption{The various CRE energy loss timescales as a function of the neutral
gas surface density ($\Sigma_{\rm gas}$). The different panels are for CREs
emitting at 0.1, 0.3, 1.4, 5 and 10 GHz.  The shaded region shows the range in
$\Sigma_{\rm gas}$ where the spectral index is observed to change (see
Figure~\ref{ant-gas}). For comparing the CRE injection timescales, we plot the
supernova surface rate density (number of supernova per pc$^2$) as dotted
lines.} 
\label{timescales} 
\end{figure*}

We have studied the variation of the non-thermal spectral index with the surface
mass density of neutral gas for four spatially resolved nearby galaxies. Our
results indicate that the local ISM gas density plays an important role in
shaping up the local radio continuum spectral behavior.  It is therefore
difficult to interpret the observed galaxy-integrated power-law nature of the
spectrum, as it averages the clumpy ISM. As pointed out earlier, after the CREs
are injected into the ISM, they are subjected to diffusion and energy loss
mechanisms that depend on the in-situ ISM density. Depending on the dominant
loss process, the injection spectrum of the CREs changes differently.
Synchrotron and inverse-Compton losses have the same effect of steepening the
spectrum.  The synchrotron timescale ($t_{\rm syn}$) for CREs emitting at a
radio frequency $\nu_{\rm syn}$ and having energy $E~[\approx7.9 (\nu_{\rm
syn}/{\rm GHz})^{1/2} (B_\perp/\mu{\rm G})^{-1/2} ~{\rm GeV}]$ is given by, 
\begin{equation} 
t_{\rm syn} = 8.35 \times 10^9 \left(\frac{E}{\rm GeV}\right)^{-1} \left(\frac{B_\perp}{\rm \mu G}
\right)^{-2}~~{\rm yr}.  
\end{equation} 
Here, $B_\perp$ is the magnetic field strength in the plane of the sky.

The timescale for inverse-Compton losses ($t_{\rm IC}$) due to scattering of
CREs by radiation field in the galaxy with energy density $U_{\rm rad}$ and
cosmic microwave background with energy density $U_{\rm CMB}$ is given by,
\begin{equation}
	t_{\rm IC} = 4.5 \times 10^8 \left(\frac{E}{\rm GeV}\right)^{-1} \left(\frac{U_{\rm rad} + U_{\rm CMB}}{10^{-12} ~{\rm erg~cm^{-3}}} \right)^{-1}~~{\rm yr}.
\end{equation}

Ionization losses flattens the spectrum towards lower radio frequencies and the
timescale ($t_{\rm ion}$) is given by \nocite{murph09}({Murphy} 2009),
\begin{equation} t_{\rm ion} = 4.1 \times 10^9 \left(\frac{\langle
n\rangle}{\rm cm^{-3}}\right)^{-1} \left(\frac{E}{\rm GeV} \right)
\left[3\ln\left(\frac{E}{\rm GeV}\right) + 42.5\right]^{-1}~~{\rm yr}.
\end{equation} Here, $\langle n\rangle$ is the average number density of
neutral gas (atomic + molecular) as seen by the CREs.

Bremsstrahlung losses does not significantly change the spectrum and its
timescale ($t_{\rm brem}$) is given by \nocite{murph09}({Murphy} 2009), 
\begin{equation} 
t_{\rm brem} = 3.96\times 10^7 \left(\frac{\langle n\rangle}{\rm cm^{-3}}\right)^{-1}~~{\rm yr}.
\end{equation}

Since the ISM of a galaxy is not smooth in its physical properties, i.e.,
magnetic field strength and gas density, different losses will dominate in
different parts of the galaxy. To compare the various energy loss scenarios we
plot these energy loss timescales as a function of $\Sigma_{\rm gas}$ in
Figure~\ref{timescales}.  As these loss timescales depend on the energy of the
CRE and thereby the radio frequency, we show these timescales at 0.1, 0.3, 1.4,
5 and 10 GHz.  Further, at a fixed frequency, the effective CR energy depends
on the magnetic field.  The top axis in Figure~\ref{timescales} shows the
energy of the CREs.  The magnetic field strength ($B$) can depend on
$\Sigma_{\rm gas}$ as $B\propto \Sigma_{\rm gas}^{1/2}$ \nocite{schle13}({Schleicher} \& {Beck} 2013). Using
the normalization\footnote{Note that the normalizations in \nocite{schle13}{Schleicher} \& {Beck} (2013)
are given in terms of the surface star-formation rate ($\Sigma_{\rm SFR}$). We
converted $\Sigma_{\rm SFR}$ to $\Sigma_{\rm gas}$ following \nocite{kenni98}{Kennicutt} (1998).}
given by \nocite{schle13}{Schleicher} \& {Beck} (2013), $B\approx 2 \times (\Sigma_{\rm gas}/{\rm M_\odot
pc^{-2}})^{1/2}~\mu$G.  Further, for the inverse-Compton losses, we considered
the $U_{\rm rad}$ to depend on $\Sigma_{\rm gas}$ as, $U_{\rm rad} = 8.33
\times 10^{-16} (\Sigma_{\rm gas}/{\rm M_\odot pc^{-2}})~{\rm erg~cm^{-3}}$,
normalized in terms of the Milky Way \nocite{schle13}({Schleicher} \& {Beck} 2013).

The average number density of total neutral (atomic+molecular) gas ($\langle
n\rangle$) was estimated from surface mass density ($\Sigma_{\rm gas}$).
Typically, the surface mass density of atomic gas as traced by H{\sc i}
emission does not exceed $\sim 10~\rm M_\odot pc^{-2}$ for normal star-forming
galaxies \nocite{leroy08, kalbe08}(see e.g., {Leroy} {et~al.} 2008; {Kalberla} \& {Dedes} 2008). We therefore assumed the
regions with $\Sigma_{\rm gas} \gtrsim 25~\rm M_\odot pc^{-2}$ to be
originating entirely from molecular gas as traced by CO. Such regions are
typically found in the inner regions of the galaxies and in spiral arms. The
regions with $\Sigma_{\rm gas} \lesssim 10~\rm M_\odot pc^{-2}$, were assumed to
be originating from atomic H{\sc i} gas, typically found in interarm regions
and outer parts of the galaxies. For regions having $\Sigma_{\rm
gas}\sim10-25~\rm M_\odot pc^{-2}$, we assumed a mixture of atomic and
molecular gas such that H{\sc i}/(H{\sc i}+H$_2$) varying linearly between 1
and 0. 

The surface gas density is converted to mid-plane number density assuming a
Milky Way type model for the vertical distribution of the atomic and molecular
gas. We used a 3-component vertical distribution for the H{\sc i}, following
\nocite{dicke90}{Dickey} \& {Lockman} (1990) and a single component for H$_2$, following \nocite{sande84}{Sanders}, {Solomon} \&  {Scoville} (1984).
The three H{\sc i} components are as follows: 1) a Gaussian component with FWHM
of 212 pc ($\sigma_1 =90.2$ pc) having mid-plane density of $n_1$, 2) a second
Gaussian component with FWHM of 530 pc ($\sigma_2=225.5$ pc)
having mid-plane density of $n_2$ and 3) an exponential component with scale
height ($l$) of 403 pc with mid-plane density $n_3$ \nocite{dicke90}({Dickey} \& {Lockman} 1990). The
components have relative ratios, $n_1:n_2:n_3 =0.395:0.107:0.064$
\nocite{dicke90}({Dickey} \& {Lockman} 1990). Thus, $\Sigma_{\rm HI}$ is modeled as,
\begin{multline} 
\Sigma_{\rm HI} = \int\limits_{-\infty}^{+\infty} \big[ n_1\exp(-z^2/2\sigma_1^2) + n_2 \exp(-z^2/2\sigma_2^2) \\+ n_3 \exp(-z/l)\big]  dz
\end{multline}

Similarly, the mid-plane number density of H$_2$ ($n_{\rm H_2}$) is computed
assuming a FWHM of 110 pc \nocite{sande84}({Sanders} {et~al.} 1984) corresponding to $\sigma_{\rm H_2}=46.8$
pc. Thus, $\Sigma_{\rm H_2}$ is given by,
\begin{equation}
\Sigma_{\rm H_2} = \int\limits_{-\infty}^{+\infty} n_{\rm H_2} \exp(-z^2/2\sigma_{\rm H_2}^2)~ dz
\end{equation}

The total mid-plane number density ($n_{\rm tot}$) is given by, $n_{\rm tot} =
n_1 + n_2 + n_3 + n_{\rm H_2}$.  We assumed the average number density of
neutral gas, $\langle n\rangle$, seen by the CREs to be half of $n_{\rm tot}$.

\begin{figure*} 
\begin{center} 
\begin{tabular}{cc}
{\mbox{\includegraphics[width=9cm,height=9cm]{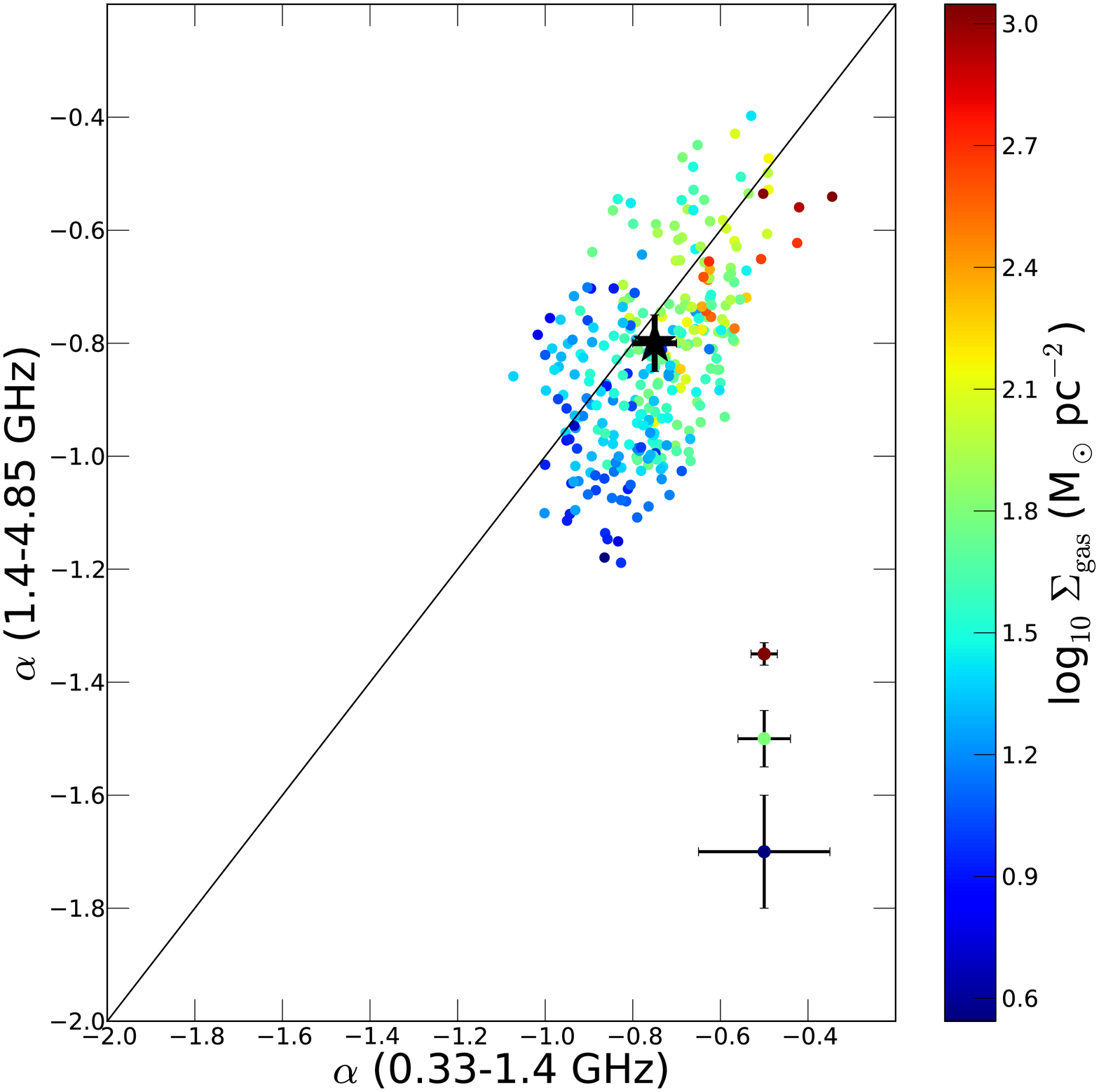}}}
{\mbox{\includegraphics[width=9cm,height=9cm]{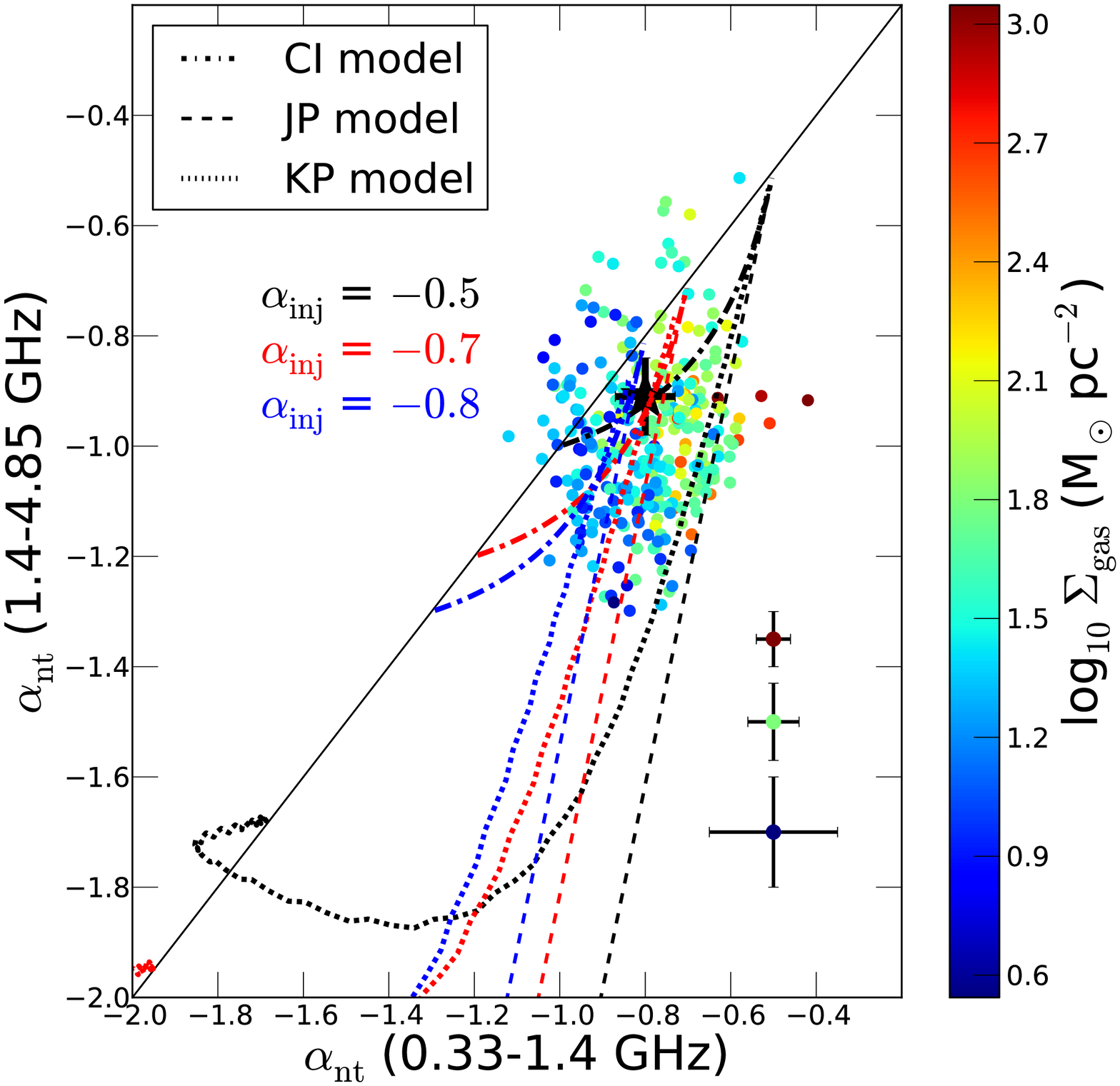}}} 
\end{tabular}
\end{center} 
\caption{{\it Left-hand panel}: Total spectral index between 1.4 and 4.85 GHz
[$\alpha_{\rm total}$(1.4--4.85 GHz)] versus spectral index between 0.33 and
1.4 GHz [$\alpha_{\rm total}$(0.33--1.4 GHz)] for the galaxy NGC 6946. {\it
Right-hand panel}: Non-thermal spectral index between 1.4 and 4.85 GHz
[$\ant$(1.4--4.85 GHz)] versus between 0.33 and 1.4 GHz [$\ant$(0.33--1.4
GHz)].  The spectral indices are measured within one synthesized beam. The star
shows the galaxy integrated spectral index. The symbols are colour coded based
on the $\log_{10}\Sigma_{\rm gas}$.  The dashed, dashed-dot and the dotted
lines shows the CI, JP and KP models, respectively, for different $\alpha_{\rm
inj}$ of $-0.5$, $-0.7$ and $-0.8$ (starting from top right). The solid
lines shows the 1:1 line. The symbols with errors represents the typical
error on the spectral index for different values of $\log_{10}\Sigma_{\rm
gas}$.}
\label{alpha_alpha} 
\end{figure*}

\subsection{CR injection}

All the CR energy loss mechanisms would be relevant if the CR injection
timescales were significantly larger. Otherwise, the loss mechanisms may not
affect the radio continuum spectrum and one would observe the spectral index
close to the injection value, $\alpha_{\rm inj}$. Under DSA, $\alpha_{\rm inj}$
depends on the Mach number ($M$) as, $\alpha_{\rm inj} = -(M^2+3)/(2M^2 -2)$
\nocite{bland87}({Blandford} \& {Eichler} 1987).  For strong shocks, $\alpha_{\rm inj}\approx -0.5$. We note
that the Mach number depends on the sound speed and thereby the gas density
($\rho$) as $M\propto \sqrt{\rho}$.  Thus, depending on the gas density,
$\alpha_{\rm inj}$ could be in the range $\sim-0.5$ to $-0.8$.  

The black dotted lines in Figure~\ref{timescales} show the injection timescales
of freshly generated CR particles estimated from the supernova rates.  The
stars that are $\gtrsim8~\rm M_\odot$ end up in Type II and Type Ib supernova,
and are responsible for acceleration of the CREs in normal galaxies
\nocite{condo92}({Condon} 1992).  Using the initial mass function (IMF) given by
\nocite{kroup01}{Kroupa} (2001), we estimated the mass fraction ($f_{\rm M}$) of stars that
could form supernovae by integrating the IMF between 8 and 150 M$_\odot$. Thus,
for a surface mass star formation rate of $\Sigma_{\rm SFR}$, the mass rate of
forming supernova stars is given by, $f_{\rm M}\times \Sigma_{\rm SFR}$.  The
mean surface supernova rate ($\nu_{\rm SN}$) is given by, $\nu_{\rm SN} =
f_{\rm M}\times \Sigma_{\rm SFR}/\langle M_{\rm SN}\rangle$.  Here, $\langle
M_{\rm SN}\rangle$ is the mean supernova mass. In our case, $f_{\rm M}/\langle
M_{\rm SN}\rangle$ is found to be 0.01.  $\Sigma_{\rm SFR}$ was estimated from
$\Sigma_{\rm gas}$ using the Kennicutt-Schmidt law \nocite{kenni98}({Kennicutt} 1998). From
Figure~\ref{timescales} we find the local CR injection timescales to be larger
than the loss timescales and hence, the freshly generated CREs do not
contribute significantly to the observed synchrotron spectrum. Except for
regions of high star-formation or recent starburst activity, an observed 
flatter spectral index of $-0.5$, close to the injection value, is
unlikely to be caused by the fresh CREs.

\subsection{Gas density and energy losses}

From Figure~\ref{timescales} it is clear that ionization and bremsstrahlung
losses dominate in regions of high gas density at lower radio frequencies 
($\lesssim 0.3$ GHz) where the synchrotron emitting CREs have less energy
($\sim 1$ GeV).  Inverse-Compton losses are mostly un-important within
galaxies, except for very low density regions with magnetic field strength
$<3.3~\mu$G.  For regions with $\Sigma_{\rm gas}\gtrsim50~\rm M_\odot pc^{-2}$,
$t_{\rm ion}$ and $t_{\rm brem}$ are lower than that of $t_{\rm syn}$
below $\sim0.3$ GHz. Hence, such regions are expected to have flatter
spectral index as is observed in Figure~\ref{ant-gas} (shown as shaded area in
Figure~\ref{timescales}). The regions with $\Sigma_{\rm gas}\gtrsim200~\rm
M_\odot pc^{-2}$ are perhaps contributed by dense molecular clouds having high
densities and magnetic field strengths. In such regions, ionization losses can
become significant at $\sim 0.3$ GHz, if the magnetic field strengths are
high, giving rise to spectral index flatter than their injection value of
$\sim-0.5$. This is seen for the galaxy NGC 5236, where regions with
$\Sigma_{\rm gas} \gtrsim 200~\rm M_\odot pc^{-2}$ are observed to have $\ant$
in the range $-0.3$ to $-0.5$. Such regions have been observed to have
relatively high magnetic field strengths \nocite{basu13}($\gtrsim 25~\mu$G; {Basu} \& {Roy} 2013).
Bremsstrahlung losses may, however, dominate significantly in moderately dense
regions with $\gtrsim100~\rm M_\odot pc^{-2}$ at 1.4 GHz.  Such regions do not
affect $\ant$ and hence, we do not expect to observe $\ant$ flatter than $-0.5$
when measured between radio frequencies above 1.4 GHz.  For ionization losses
to dominate over synchrotron losses at higher radio frequencies (between
$\sim1.4-5$ GHz), an extremely high surface gas density of $\gtrsim1000~\rm
M_\odot pc^{-2}$ is required. However, in such regions, the injection timescale
of CREs is comparable to the loss timescales within a factor of $\sim2$.

Regions with $\Sigma_{\rm gas} \lesssim 50~\rm M_\odot pc^{-2}$, are 
observed to have steep $\ant$ (see Figure~\ref{ant-gas}). From
Figure~\ref{timescales}, we find that such regions are dominated by synchrotron
and/or inverse-Compton losses at 1.4 GHz and hence perhaps causes steepening of
$\ant$.  However, note that the CRE injection timescale approaches few times
$10^9$ yr in regions of even lower $\Sigma_{\rm gas}$ ($\lesssim10~\rm M_\odot
pc^{-2}$), typically found in the interarm regions and outer parts of the
galaxies. The CRE population in such regions are dominated by propagation from
star-forming regions in adjacent arms. Depending on the propagation mechanism,
i.e., simple diffusion or streaming instability at Alfv{\' e}n velocity, the
CREs at 0.33 GHz propagate 1.4 to 2 times longer distances than those at 1.4
GHz \nocite{basu13}(see {Basu} \& {Roy} 2013). This could also give rise to the observed steepening
of $\ant$ between the two frequencies in such regions.

To test the scenario of the expected differences in spatially resolved $\ant$,
we studied $\ant$ between a low frequency pair (0.3 and 1.4 GHz) and
between a high frequency pair (1.4 and 4.85 GHz). The galaxy NGC 6946 has the
best resolution spectral index map in our sample.  NGC 6946 was observed at
4.85 GHz using the VLA in D-array configuration and a combination with a single
dish Effelsberg observation was done to ensure no missing flux density
\nocite{beck07}({Beck} 2007). In Figure~\ref{alpha_alpha} (left-hand panel) we plot the total
radio emission spectral index between 1.4 and 4.85 GHz [$\alpha$(1.4--4.85
GHz)] versus the spectral index between 0.33 and 1.4 GHz [$\alpha$(0.33--1.4
GHz)]. The total emission spectral index estimated between the two sets of well
separated radio frequencies follow the 1:1 line closely. The symbols are colour
coded based on $\log_{10} \Sigma_{\rm gas}$. In the high gas density regions,
typically in the arms and inner regions of the galaxy, the spectral indices are
flatter. However, these values could be affected due to the presence of high
thermal emission.

In Figure~\ref{alpha_alpha} (right-hand panel), we plot the non-thermal spectral
index estimated between 1.4 and 4.85 GHz [$\ant$(1.4--4.85 GHz)] and between
0.33 and 1.4 GHz [$\ant$(0.33--1.4 GHz)] (see Section~2). In general,
$\ant$(1.4--4.85 GHz) is found to be steeper than $\ant$(0.33--1.4 GHz). It is
steeper by up to $-0.5$ in regions of lower gas surface density (bluer points),
i.e., the interarm regions and outer parts of the galaxy. This is consistent
with the scenario that such regions are dominated by synchrotron and/or
inverse-Compton losses resulting in steeper spectral indices at higher
frequencies.  

We compare the expected variation of $\ant$ measured between the two pairs of
radio frequencies using the different models of energy loss, i.e., the continuous
injection (CI), Jaffe-Perola (JP) and Kardashev-Pacholczyk (KP) models.  We
generated synthetic radio spectra for each of the models with varying break
frequencies.  In Figure~\ref{alpha_alpha} (right-hand panel) we plot the
trajectory of $\ant$ for a constant injection spectral index of $\alpha_{\rm
inj}$.  The CI model is indicated by the dashed-dot lines, the JP model is
indicated by dashed lines and the KP model is shown as the dotted lines.  It is
clear that the CI model cannot give rise to the observed steepening of the
spectrum between the two pairs of radio frequencies of our study.  This
is also indicated in Figure~\ref{timescales}, which shows that the injection
timescales are larger than the loss timescales. The JP and KP models reproduce
the observed steepening at higher frequency better. The maximum steeping
observed in our case corresponds to a break frequency of $\sim4$ GHz for the JP
model and $\sim2.5$ GHz for the KP model. Detailed modelling of the spatially
resolved radio continuum spectra would require high quality maps at several
frequencies covering a wide range.

Moreover, we note that the regions of high gas density are also the regions
with dominant ionization and/or bremsstrahlung losses. Such losses have the
effect of flattening $\ant$ at lower frequencies as seen in
Figure~\ref{ant-gas}. As is also evident from Figure~\ref{timescales}, at
regions of high surface gas density ($\gtrsim200~\rm M_\odot pc^{-2}$),
ionization losses should start dominating at $\sim0.3$ GHz.  Such regions
are observed in the galaxy NGC 5236 and are likely to be giant molecular
clouds. We observe $\ant$ to be flatter than the typical injection values of
$-0.5$.  However, these losses will not affect $\ant$ at higher radio
frequencies. We therefore observe the values of $\ant$(1.4--4.85 GHz) to be
well mixed in all regions with no clear trend\footnote{Note the systematic
trend of density variation (red to orange to yellow to blue points) as one
moves from right to left along $x-$axis.  No such trend is observed while
moving from top to bottom along $y-$axis.} with gas density.

Due to the smooth transition of the values of $\ant$ at lower and higher
frequencies in the spatially resolved case, the galaxy integrated case does not
capture this scenario well, i.e., no significant change in spectral index
should be expected.  The stars in both the panels in Figure~\ref{alpha_alpha}
shows the galaxy integrated spectral index. The spectral index for total radio
emission is observed to be close to the 1:1 line. For the non-thermal emission
(right-hand panel), the integrated spectrum is still observed to remain close
to the 1:1 line.  Within errors, we observe both total and non-thermal spectral
index to remain close to $\sim-0.8$ for both pairs of frequencies.  Hence, we
do not see any significant change of the slope for the galaxy integrated
spectrum.  Moreover, the dense regions that shows evidence of ionization
losses, cover only a small fraction ($\lesssim5$ percent) of the galaxy. Thus,
the galaxy integrated spectrum may not show such a behavior.  Our study
comprehensively shows that different loss mechanisms are dominant in different
parts of the galaxies. It is difficult to single out any one of the energy loss
mechanisms while interpreting the galaxy integrated spectrum.  Such studies,
hence do not often show significant deviation from a power-law behavior for a
wide range of radio frequencies.

Data on more galaxies are required to see such clear trends. LOFAR and EVLA are
ideal instruments to do such studies robustly at lower ($<150$ MHz) and higher
($>1.4$ GHz) radio frequencies, respectively. Complementary data from the GMRT
covering the intermediate frequency range ($150-610$ MHz) could help us model
the spatially resolved spectral behavior of continuum emission in detail. High
resolution CO observations from the ALMA could facilitate measuring the
distribution of the molecular gas at unprecedented detail.

\section{Conclusions}

We have studied the variation of spatially resolved non-thermal spectral index
($\ant$) as a function of the total neutral gas surface mass density
($\Sigma_{\rm gas}$) for four nearby galaxies. The galaxies in our sample are,
NGC 4736, NGC 5055, NGC 5236 and NGC 6946. In this study, we have measured the
$\ant$ and $\Sigma_{\rm gas}$ at sub-kpc linear scales except for the galaxy
NGC 5236. The spectral index were measured between 0.33 and 1.4 GHz. We present
our conclusions here.

\begin{enumerate}[(i)]

\item The local ISM gas density plays an important role in shaping up the 
radio continuum spectrum.  We find the $\ant$ to remain comparatively
flat ($\gtrsim -0.7$) in the regions of high gas density ($\Sigma_{\rm
gas}\gtrsim50~\rm M_\odot pc^{-2}$) and steepens sharply ($<-1$) in regions of
low gas density ($\Sigma_{\rm gas}\lesssim50~\rm M_\odot pc^{-2}$).

\item The observed flattening is unlikely to be caused by free--free absorption at
low frequency such as 0.33 GHz. We find that in the regions of high gas density
($\gtrsim 50~\rm M_\odot pc^{-2}$), bremsstrahlung and ionization losses could 
flatten the $\ant$ at radio frequencies below $\sim 0.33$ GHz.

\item Synchrotron losses dominates at higher radio frequencies ($\gtrsim 5$ GHz)
in regions having moderate densities ($\Sigma_{\rm gas} \lesssim 100~\rm
M_\odot pc^{-2}$). In denser regions, bremsstrahlung losses are expected
to dominate.  Inverse-Compton losses are mostly unimportant in galaxies except
for regions of very low gas densities having magnetic field strengths
$<3.3~\mu$G.

\item For the galaxy NGC 6946, we study the variation of $\ant$ measured
between 0.33 and 1.4 GHz and between 1.4 and 4.85 GHz. We find that $\ant$ between
the higher frequency pair is steeper than that between the lower frequency pair
indicating synchrotron or inverse-Compton losses at 4.85 GHz. The observed
steepening cannot be caused by continuous injection of CR particles and requires
single shot particle injection.

\item Due to the clumpy nature of the ISM, these local CRE energy losses may not
show indications in the overall galaxy-integrated spectrum. Thus, the
galaxy-integrated spectrum may remain a power-law for a wide range of radio 
frequencies.

\end{enumerate}

\section*{Acknowledgments}

We thank the referee, Joshua Marvil, for the helpful comments and
suggestions that improved the presentation of the paper. We thank Nirupam Roy
for extensive discussions and critically going through the manuscript that
considerably improved the presentation of the paper. 
This work has made use of
HERACLES, `The HERA CO-Line Extragalactic Survey’ (Leroy et al.  2009). This
work has made use of THINGS, `The H{\sc i} Nearby Galaxy Survey’ (Walter et al.
2008).




\bsp

\label{lastpage}

\end{document}